\documentclass[aps,prc,preprint,tightenlines,floatfix,groupedaddress,
nofootinbib,showpacs,preprintnumbers,amsmath,amssymb,superscriptaddress]{revtex4}
\usepackage{graphicx}
\usepackage{dcolumn}
\usepackage{mathrsfs}
\usepackage{bm}

\usepackage{graphicx}
\usepackage{dcolumn}
\usepackage{bm}
\usepackage[usenames]{color}

\begin{document}

\title{Proton fraction in the inner neutron-star crust }

\author{J. Piekarewicz}
\email{jpiekarewicz@fsu.edu}
\affiliation{Department of Physics, Florida State University, 
                  Tallahassee, Florida 32306, USA}
\author{G. Toledo S\'anchez}
\email{toledo@fisica.unam.mx}
\affiliation{Departamento de Fisica Te\'orica, Instituto de Fisica,
                  Universidad Nacional Aut\'onoma de M\'exico, 
                  A.P. 20-364, M\'exico 01000 D.F., M\'exico}
\date{\today}

\begin{abstract}
 Monte Carlo simulations of neutron-rich matter of relevance to 
 the inner neutron-star crust are performed for a system of
 $A\!=\!5,000$ nucleons. To determine the proton fraction in the 
 inner crust, numerical simulations are carried out for a variety of
 densities and proton fractions. We conclude---as others have 
 before us using different techniques---that the proton fraction 
 in the inner stellar crust is very small. Given that the purported 
 {\sl ``nuclear pasta''} phase in stellar crusts develops as a 
 consequence of the long-range Coulomb interaction among 
 protons, we question whether pasta formation is possible in 
 such proton-poor environments. To answer this question, we 
 search for physical observables sensitive to the transition between 
 spherical nuclei and exotic pasta structures. Of particular relevance 
 is the static structure factor $S(k)$---an observable sensitive to 
 density fluctuations. However, no dramatic behavior was observed 
 in $S(k)$. We regard the identification of physical observables 
 sensitive to the existence---or lack-thereof---of a pasta phase 
 in proton-poor environments as an open problem of critical 
 importance.
\end{abstract}
\maketitle

\section{Introduction}
\label{Introduction}

Neutron stars are unique laboratories for the study of matter under
extreme conditions of density and isospin 
asymmetry~\cite{Weber:1999,Glendenning:2000}. Indeed, the
conditions in the interior of neutron stars are so extreme that they
are unattainable in terrestrial laboratories. Thus, neutron stars---and 
the exotic phases within---owe their existence to the presence of 
enormous gravitational fields. To maintain hydrostatic equilibrium 
throughout the star, these enormous gravitational fields must be 
balanced by the pressure support of its underlying constituents. 
This results in an enormous dynamic range of pressures and 
densities that enables one to probe the equation of state (EOS)
far away from the equilibrium density of 
normal nuclei~\cite{Piekarewicz:2009zz}.

Neutron stars contain a non-uniform crust above the uniform liquid
mantle (or stellar core). The core is comprised of a uniform assembly of
neutrons, protons, electrons, and muons packed to densities that may
exceed that of normal nuclei by up to an order of magnitude.  The
highest density attained in the core depends critically on the
equation of state of neutron-rich matter which at those high densities
is presently poorly constrained. The core accounts for almost all of
the mass and most of the size of the neutron star. However, at
densities of about half of normal nuclear density, the uniform core
becomes unstable against cluster formation. At these ``low'' densities 
the average inter-nucleon separation increases to such an extent that it
becomes energetically favorable for the system to segregate into
regions of normal density (nuclear clusters) and regions of low
density (dilute neutron vapor).  Such a clustering instability signals
the transition from the uniform liquid core to the non-uniform
crust.  Note, however, that the precise value of the
crust-to-core transition density is presently unknown, as it is
sensitive to the poorly constrained density dependence of the 
symmetry energy~\cite{Horowitz:2000xj}.

The solid crust is divided into an outer and an inner
region. In particular, the outer crust spans a region of about seven
orders of magnitude in density (from about $10^{4}{\rm g/cm^{3}}$ to
$4\times 10^{11}{\rm
g/cm^{3}}$~\cite{Baym:1971pw,Ruester:2005fm,RocaMaza:2008ja}).
Structurally, the outer crust is comprised of a Coulomb lattice of
neutron-rich nuclei embedded in a uniform electron gas. As the density
increases---and given that the electronic Fermi energy increases
rapidly with density---it becomes energetically favorable for
electrons to capture into protons. This results in the formation of
Coulomb crystals of progressively more neutron-rich nuclei.
Eventually, the neutron-proton asymmetry becomes too large for the
nuclei to absorb any more neutrons and the excess neutrons go into the
formation of a dilute neutron vapor; this signals the transition from
the outer to the inner crust.

In this contribution we are interested in modeling the structure of
the inner crust.  The outer-to-inner crust transition density is
predicted to occur at about $4\times 10^{11}{\rm
g/cm^{3}}$~\cite{Baym:1971pw,Ruester:2005fm,RocaMaza:2008ja}.  
At this---{\sl neutron-drip}---density the neutron-rich nucleus
(${}^{118}$Kr) that comprises the crystalline lattice is unable to
retain any more neutrons. Thus, the top layer of the inner crust
consists of a Coulomb crystal of neutron-rich nuclei immersed in 
a uniform electron gas and a dilute---likely superfluid---neutron 
vapor. In contrast, at the bottom layer of the inner crust the density 
has become high enough (of the order of $10^{14}{\rm g/cm^{3}}$) 
to {\sl ``melt''} the crystal into a uniform mixture of neutrons,
protons, and electrons. Yet the transition from the highly-ordered
crystal to the uniform liquid is both interesting and complex. This
is because distance scales that were well separated in both the
crystalline phase (where the long-range Coulomb interaction 
dominates) and in the uniform phase (where the short-range strong 
interaction dominates) become comparable. This unique situation 
gives rise to {\sl ``Coulomb frustration''}.  Frustration, a
phenomenon characterized by the existence of a very large number of
low-energy configurations, emerges from the impossibility to
simultaneously minimize all {\sl ``elementary''} interactions in the
system. Whereas protons are correlated at short distances by
attractive strong interactions, they are anti-correlated at large
distances because of the Coulomb repulsion. Whenever these short and
large distance scales are well separated (as in the outer crust)
protons bind into nuclei that are then segregated in a crystal
lattice. However, as these length scales become comparable---at
densities of about $10^{13}\!-\!10^{14}$~g/cm$^3$ as in the inner
crust---competition among the elementary interactions results in the
formation of complex topological structures collectively referred to
as {\sl ``nuclear pasta''}. Given that these complex structures are 
very close in energy, it has been speculated that the transition from
the highly ordered crystal of spherical nuclei to the uniform phase
must proceed through a series of changes in the dimensionality and
topology of these
structures~\cite{Ravenhall:1983uh,Hashimoto:1984}. Moreover, due to
the preponderance of low-energy states, frustrated systems display
an interesting and unique low-energy dynamics.

Interestingly enough, a seemingly unrelated condensed-matter
problem---{\sl the strongly-correlated electron gas}---appears to be
connected to the nuclear pasta. In the case of the electron
gas, one aims to characterize the transition from the low-density
Wigner crystal (where the Coulomb potential dominates) to the uniform
high-density Fermi liquid (where the kinetic energy
dominates)~\cite{Fetter:1971}.  It has been argued that instead of the
expected first order phase transition, the transition is mediated by
the emergence of {\sl ``microemulsions''}, namely, exotic
(``pasta-like'') structures.  Indeed, it has been shown that in {\sl
two-spatial} dimensions first-order phase transitions are forbidden in
the presence of long-range ({\sl e.g.,} Coulomb)
forces~\cite{Jamei:2005}. One should note that no generalization of
this theorem to three dimensions exists. Hence, although the existence
of pasta phases in both core-collapse supernovae and neutron stars may
be plausible~\cite{Ravenhall:1983uh,Hashimoto:1984}, there is no
guarantee that they exist. Actually, in a recent publication Oyamatsu
and Iida have shown that pasta formation may not be universal and that
its existence (or lack-thereof) is intimately related to the density
dependence of the symmetry energy~\cite{Oyamatsu:2006vd}.  In
particular, the authors concluded that pasta formation requires models
with a soft symmetry energy, namely, ones that increase slowly with
density.

Although the complex dynamics of the electron gas may shed light on
the possible emergence of pasta phases in the stellar crust, an
essential difference between these two systems remains. Whereas the
constituents of the electron gas ({\sl i.e.,} electrons) are all
electrically charged and thus experience long-range forces, the
neutrons in the crust interact solely via short-range forces.  Thus, a
large neutron fraction in the inner crust could hinder the formation
of the nuclear pasta. Indeed, pure neutron matter does not cluster.
So given that the proton fraction in the crust is highly sensitive to
the density dependence of the symmetry energy, the emergence of pasta
phases involves a delicate interplay between: (i) the symmetry energy
(which controls the proton fraction), (ii) the surface tension (which
favors spherical nuclei), and (iii) the Coulomb interaction (which
favors deformation). It is this interesting and unique problem that
we propose to study here via numerical simulations.

Monte-Carlo simulations will be performed directly in terms of the
nucleon constituents. Although we have already performed simulations
of this kind~\cite{Horowitz:2004yf,Horowitz:2004pv,Horowitz:2005zb},
in the present manuscript we aim to improve them in two critical
areas.  First, in our previous work a screened Coulomb interaction
between the protons was assumed with a screening length fixed at
$10$~fm.  Although it was argued that no major qualitative changes are
expected from such an approximation, in this contribution we evaluate
the Coulomb interaction exactly via an Ewald summation.  For
completeness, an appendix has been included where a detailed
derivation of the Ewald summation is presented for a system of protons
embedded in a uniform electron gas. Second, our earlier simulations
were all performed at the single proton fraction of $Y_{p}\!=\!0.2$.
This value was chosen as it is intermediate between the large values
attained in core-collapse supernovae and the low proton fractions
characteristic of the inner crust of cold---fully catalyzed---neutron
stars.  Instead, in this work we will simulate neutron-rich matter for
a variety of densities and proton fractions, albeit for a relatively
small number of nucleons ($A\!=\!5,000$). In this way, the optimum
proton fraction will be determined by imposing $\beta$-equilibrium.

We should note that for the large proton fractions characteristic of
core-collapse supernovae ($Y_{p}\!=\!0.3$--$0.5$) there appears to be
agreement that the transition from the ordered Coulomb crystal to the
uniform phase must proceed via intermediate pasta phases (such as
rods, slabs, tubes, {\sl etc.}). This agreement has been reached
whether one uses numerical
simulations~\cite{Horowitz:2004yf,Horowitz:2004pv,
Horowitz:2005zb,Watanabe:2003xu,Watanabe:2004tr, Watanabe:2009vi} or
mean-field approximations with a variety of effective
interactions~\cite{Maruyama:2005vb,Avancini:2008zz,Avancini:2008kg,
Newton:2009zz,Shen:2011kr}.  What is unclear, however, is whether such
exotic pasta shapes will still develop in the proton-poor environment
of the inner stellar crust. For example, mean-field models that impose
$\beta$-equilibrium predict proton fractions at densities of relevance
to the inner crust of only a few
percent~\cite{Maruyama:2005vb,Avancini:2008zz,Avancini:2008kg,
Shen:2011kr}. To our knowledge, no numerical simulations have been
carried out with proton fractions less than
$Y_{p}\!=\!0.1$~\cite{Watanabe:2003xu}. Thus, our manuscript revolves
around the following three fundamental questions: (a) Do numerical
simulations support the very low proton fractions predicted by
mean-field calculations? (b) Does the transition from a highly-ordered
crystal of spherical nuclei to a uniform Fermi liquid must proceed via 
intermediate pasta phases even when the proton fraction is very small?  
(c) Can one identify a set of robust physical observables that are
sensitive to the formation of the nuclear pasta?

We have organized the manuscript as follows. In Sec.~\ref{Formalism}
we introduce the model and develop the formalism necessary to carry
out the Monte-Carlo simulations. Results are presented in
Sec.~\ref{Results} for the potential energy as a function of density
and proton fraction. The optimal proton fraction is then determined by
minimizing the total energy per nucleon of the system.  The
pair-correlation function (for both protons and neutrons) and the
corresponding static-structure factor are computed with the aim of
identifying sensitive observables to the formation of the nuclear
pasta. Our conclusions and suggestions for future work are presented
in Sec.~\ref{Conclusions}.

\section{Formalism}
\label{Formalism}

We start this section by reviewing the semi-classical model that while
simple, contains the essential physics of {\sl ``Coulomb
frustration''}, namely, competing interactions consisting of a
short-range nuclear attraction and a long-range Coulomb
repulsion~\cite{Horowitz:2004yf}.  We stress that the long-range
Coulomb interaction among the protons represents the critical
ingredient for pasta formation. Hence the importance of establishing
whether the proton fraction in the inner stellar crust is large enough
to help (or hinder) the formation of the exotic pasta structures. A
variety of static observables will be computed using Monte-Carlo
simulations having neutrons, protons, and electrons as the main
constituents. As the system is simulated directly in terms of its
basic constituents, there is no need to assume {\sl a-priori} any 
specific nuclear shape, such as rods, slabs, tubes, {\sl etc.}

\begin{figure}[htb]
\vspace{-0.05in}
\includegraphics[width=0.5\columnwidth,angle=0]{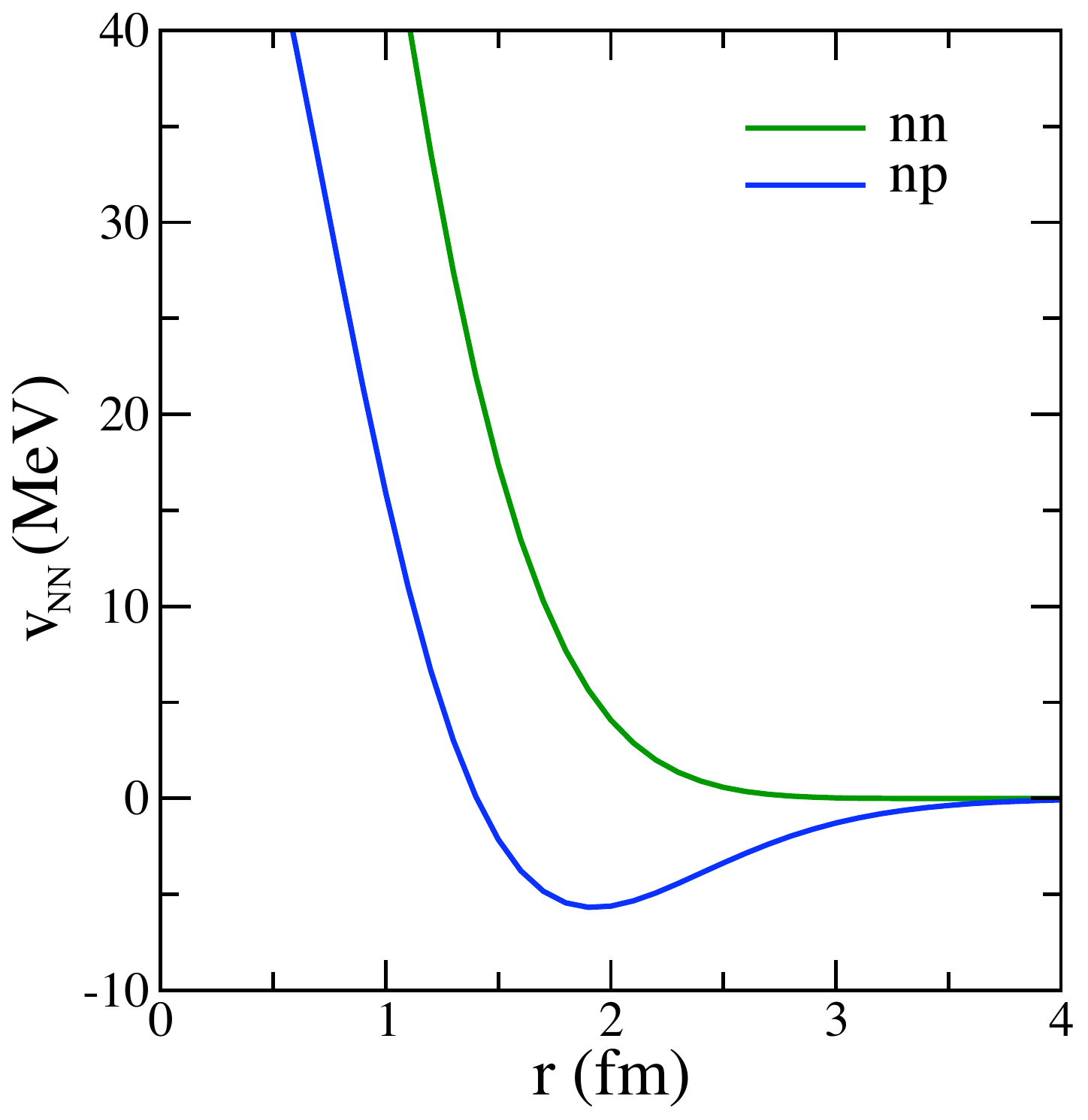}
\caption{(Color online) Nucleon-nucleon potential employed 
in our semi-classical simulations~\cite{Horowitz:2004yf}.}
\label{Fig1}
\end{figure}

The total potential energy of the system includes both short-range 
nuclear and long-range Coulomb contributions. That is,
\begin{equation}
 V({\bf r}_{1},\ldots,{\bf r}_{A}) =
  V_{\rm Nucl}({\bf r}_{1},\ldots,{\bf r}_{A}) +
  V_{\rm Coul}({\bf r}_{1},\ldots,{\bf r}_{A}) \;.
\label{VTotal}
\end{equation}
The nuclear potential is assumed to consist of a sum of 
``elementary'' two-body interactions of the following 
form:
\begin{equation}
  V_{\rm Nucl}({\bf r}_{1},\ldots,{\bf r}_{A}) =
  \sum_{i<j=1}^{A} v_{{}_{NN}}(r_{ij}) =
  \sum_{i<j=1}^{A} \Big[a e^{-r_{ij}^{2}/\Lambda} +
       \Big(b+c\,\tau_{i}\tau_{j}\Big)
	e^{-r_{ij}^{2}/2\Lambda}\Big]\;,
 \label{v}
\end{equation}
where the inter-particle distance has been denoted by
$r_{ij}\!\equiv\!|{\bf r}_i\!-\!{\bf r}_j|$ and the isospin of the
nucleon as $\tau$---with $\tau\!=\!1$ for a proton and $\tau\!=\!-\!1$
for a neutron. The model parameters ($a$, $b$, $c$, and $\Lambda$)
were calibrated in Ref.~\cite{Horowitz:2004yf} and are given by the
following values: $a\!=\!110$~MeV, $b\!=\!-26$~MeV, $c\!=\!24$~MeV,
and $\Lambda\!=\!1.25~{\rm fm}^{2}$. Although not accurately
calibrated, this set of parameters were fitted to several ground-state
properties~\cite{Horowitz:2004yf}. In particular, Monte Carlo
simulations using this simple interaction are able to account for the
saturation of symmetric nuclear matter while precluding the binding of
pure neutron matter at all densities~\cite{Horowitz:2004yf}.  Note,
however, that given that such {\sl classical} simulations are unable
to reproduce the momentum distribution of a {\sl quantum} Fermi gas,
the effective $NN$ interaction must be properly adjusted to compensate
for this shortcoming.  In particular---unlike a realistic
neutron-neutron ($nn$) interaction that is attractive at intermediate
distances--- the effective $nn$ interaction adopted here is
(essentially) repulsive at all distances; see Fig.~\ref{Fig1}.

For the Coulomb interaction we assume that the protons are immersed in
a uniform neutralizing electron background. Note that at the densities of 
relevance to the stellar crust, the electrons behave to a good approximation 
as a non-interacting free Fermi gas. For such a system the Coulomb energy 
may be written in terms of the electrostatic potential as follows:
\begin{equation}
  V_{Coul}({\bf r}_{1},\ldots,{\bf r}_{Z}) = 
\frac{1}{2}e\sum_{i=1}^{Z} \Phi({\bf r}_{i}) -
 \frac{eZ}{2V}\int_{V} \Phi({\bf r}) d^{3}r \;.
 \label{VCoulomb}
\end{equation}
Here $Z$ is the number of protons (and neutralizing electrons), $V$ is
the simulation volume, and the electrostatic potential $\Phi({\bf r})$
satisfies Poisson's equation. Given that numerical simulations must
involve a finite number of particles, we rely on periodic boundary
conditions in an attempt to minimize {\sl finite-size} effects. For
the case of the short-range nuclear interaction, the implementation of
the {\sl minimum image convention} is both simple and accurate: a
given particle in the system interacts only with the closest image of
all remaining
particles~\cite{Allen:1987,Frenkel:1996,Vesely:2001}. Indeed, the
short-range character of the $NN$ force guarantees that the
interactions with all images that are farther away will be
exponentially suppressed. This prescription, however, is not valid for
the long-range Coulomb potential. In this case a given particle in $V$
interacts not only will all remaining particles in the simulation
volume but, in addition, with {\sl all} periodic images.  As such, the
source of the electrostatic potential $\Phi({\bf r})$ in Poisson's
equation consists of all protons in the simulation volume $V$, all
their periodic images, and the uniform electron background. That is,
\begin{equation}
 \rho({\bf r}) = e\sum_{i=1}^{Z}\sum_{{\bf n}}
 \delta\Big({\bf r}-({\bf r}_{i}-{\bf n}L)\Big) - e\frac{Z}{V} \;,
\label{RhoCharge}
\end{equation}
where ${\bf n}=(n_{x},n_{y},n_{z})$ is a triplet of integers and
$L\!\equiv\!V^{1/3}$. 

A method that computes accurately and efficiently the Coulomb
potential is based on the (90 year old!) Ewald
summation~\cite{Ewald:1921}.  The basic idea behind the Ewald sum is
to add and subtract $Z$ individual {\sl smeared} charges at the exact
location of each (point) proton.  The role of each negative charge is
to screen the corresponding (point) proton charge over a distance of
the order of the smearing parameter $a$. As long as $a$ is
significantly smaller than the box length $L$, the resulting
(screened) two-body potential will become short ranged and thus
amenable to be treated using the minimum-image convention.  What
remains then is a periodic system of {\sl smeared} positive charges
together with the neutralizing electron background. In configuration
space this remaining {\sl long-range} contribution is slowly
convergent. The great merit of the Ewald sum is that the long-range
contribution can be made to converge rapidly if evaluated in momentum
space, {\sl i.e.,} as a Fourier sum.  Indeed, the Fourier sum is
rapidly convergent because momentum ({${\bf k}$) modes satisfying
$|{\bf k}|a\!\gg\!1$ make a negligible contribution to the sum. Hence,
by suitably tuning the value of the smearing parameter $a$, the
evaluation of the Coulomb potential may be written in terms of two
rapidly convergent sums; one in configuration space and one in
momentum space.  The derivation and implementation of the Ewald sum is
now part of the standard
literature~\cite{Allen:1987,Frenkel:1996,Toukmaji_199673,Vesely:2001}.  
Yet, for both convenience and completeness, we have provided in the
appendix a detailed derivation of the Ewald formula for a system of
$Z$ protons immersed in a neutralizing and uniform electron
background. Here we only summarize the essential results.

Assuming $Z$ protons confined to a simulation volume $V\!=\!L^{3}$
and immersed in a uniform neutralizing electron background, the 
Coulomb potential may be written in the following form:
\begin{equation}
  V_{\rm Coul}({\bf r}_{1},\ldots,{\bf r}_{Z}) = 
  \left(\frac{e^{2}}{L}\right)  
  U_{\rm Coul}({\bf s}_{1},\ldots,{\bf s}_{Z}) \;.
 \label{UCoulDef}
\end{equation}
This structure reveals that the Coulomb potential is an interaction
with no intrinsic scale. That is, once a dimensionful parameter has
been identified---$v_{0}\!\equiv\!e^{2}/L$ in the present case---the
dimensionless Coulomb potential $U$ depends exclusively on the 
scaled coordinates ${\bf s}_{i}\!\equiv\!{\bf r}_{i}/L$. As mentioned
above, the Ewald construction relies on the introduction of an
artificial smearing parameter $a$ that naturally divides the Coulomb
potential into two contributions: a short-range contribution that
converges rapidly in configuration space and a long-range one that
converges rapidly in momentum space. That is,
\begin{equation}
  U_{\rm Coul}({\bf s}_{1},\ldots,{\bf s}_{Z}) =
   \frac{1}{2}\sum_{i\ne j}
   \Big[u_{\rm sr}({\bf s}_{ij}) +
          u_{\rm lr}({\bf s}_{ij}) \Big] + U_{0} \;.\label{UCoulEwald1}
\end{equation}
Here the---{\sl two-body}---short- and long-range contributions 
are given by
\begin{subequations}
 \begin{align}
  u_{\rm sr}({\bf s}) & = \frac{{\rm erfc}(s/s_{0})}{s} \;,
    \label{uShort} \\
  u_{\rm lr}({\bf s}) & =\sum_{{\bf l}\ne 0} w(l) 
   \exp(-2\pi i{\bf l}\cdot{\bf s}) = \sum_{{\bf l}\ne 0} 
   \frac{\exp(-\pi^{2}s_{0}^{2}\,l^{2})}{\pi l^{2}}
   \exp(-2\pi i{\bf l}\cdot{\bf s}) \;,
    \label{uLong} 
 \end{align}
 \label{uDefs}
\end{subequations}
where $s_{0}\!\equiv\!a/L$ is the dimensionless smearing parameter  
and $U_{0}$ is an overall constant:
\begin{equation}
 U_{0} = \frac{Z}{2}u_{\rm lr}(0) 
          - \frac{Z}{\sqrt{\pi}s_{0}} 
          - \frac{\pi s_{0}^{2}Z^{2}}{2} \;.
 \label{ConstantU}
\end{equation}
The short-range contribution [Eq.~(\ref{uShort})] represents a
(gaussianly) screened two-body Coulomb potential. Indeed, the
characteristic $1/s$ fall-off of the Coulomb potential is modified by
the complement of the error function which falls rapidly to zero for
inter-particle separations significantly larger than the screening
length $s_{0}$. The long-range contribution to the Coulomb potential
is customarily written in momentum space as a sum that involves the
square of the charge form factor times a suitably modified Coulomb
propagator in momentum space [see Eq.~(\ref{VCoulLR1})]. Rapid
convergence of the momentum sum is also achieved through the
introduction of the smearing parameter [see Eq.~(\ref{uLong})].
However, the evaluation of the momentum sum often remains the most
numerically demanding part of the computation, so various efficient
methods have been designed for the purpose of expediting the Fourier
sum~\cite{Frenkel:1996}. In our case we will settle for a relatively
simple approach that ultimately leads to the form of $u_{\rm lr}({\bf
s})$ given above (see appendix). Not surprisingly, this form also
involves a numerically demanding Fourier sum. Yet, the advantage of
the present method is that $u_{\rm lr}({\bf s})$ may be precomputed in
a fine mesh before the start of the simulation. Hence, during the
actual simulation the evaluation of $u_{\rm lr}({\bf s})$ is
implemented via a look-up table and a simple interpolation
scheme~\cite{Toukmaji_199673}.  This approach results in significant
savings in processing time with little compromise in accuracy. In the
present contribution we have selected a smearing parameter that is
significantly smaller than the box length, namely,
$s_{0}\!=\!0.12$.

\section{Results}
\label{Results}

In this section we present results for the potential energy per
nucleon as a function of density and proton fraction as obtained from
our Monte Carlo simulations. In an effort to simulate the
quantum-mechanical {\sl zero-point motion} of the particles, the
simulations were carried out at a fixed temperature of
$T\!=\!1$~MeV~\cite{Horowitz:2004yf,Horowitz:2004pv,Horowitz:2005zb}.
Moreover, in all simulations the number of baryons was fixed at
$A\!=\!5,000$ with the proton fraction varying from $Y_{p}\!=\!0$ to
$Y_{p}\!=\!0.25$; this is contrast to our earlier simulations that assumed
a fixed value of $Y_{p}\!=\!0.2$. Having fixed the number of baryons, the
proton fraction, the temperature, and the density of the system, a
Metropolis algorithm was used to generate a total of 3,000 Monte Carlo
sweeps---with each sweep consisting of $A\!=\!5,000$ 
particle moves. The first 2,500 sweeps were used to thermalize the
system and the last 500 to compute various physical observables. This
was done {\sl ``off-line''} by first saving the coordinates of all
$5,000$ particles to file (for each of the final 500 sweeps) and
then using a post-processor to compute the observables. We focus
initially on the energy per particle as a function of density and
proton fraction as we aim to estimate whether the proton fraction 
in the bottom layers of the inner stellar crust is small enough to
hinder the formation of the pasta.

\begin{figure}[htb]
\vspace{-0.05in}
\includegraphics[width=0.6\columnwidth,angle=0]{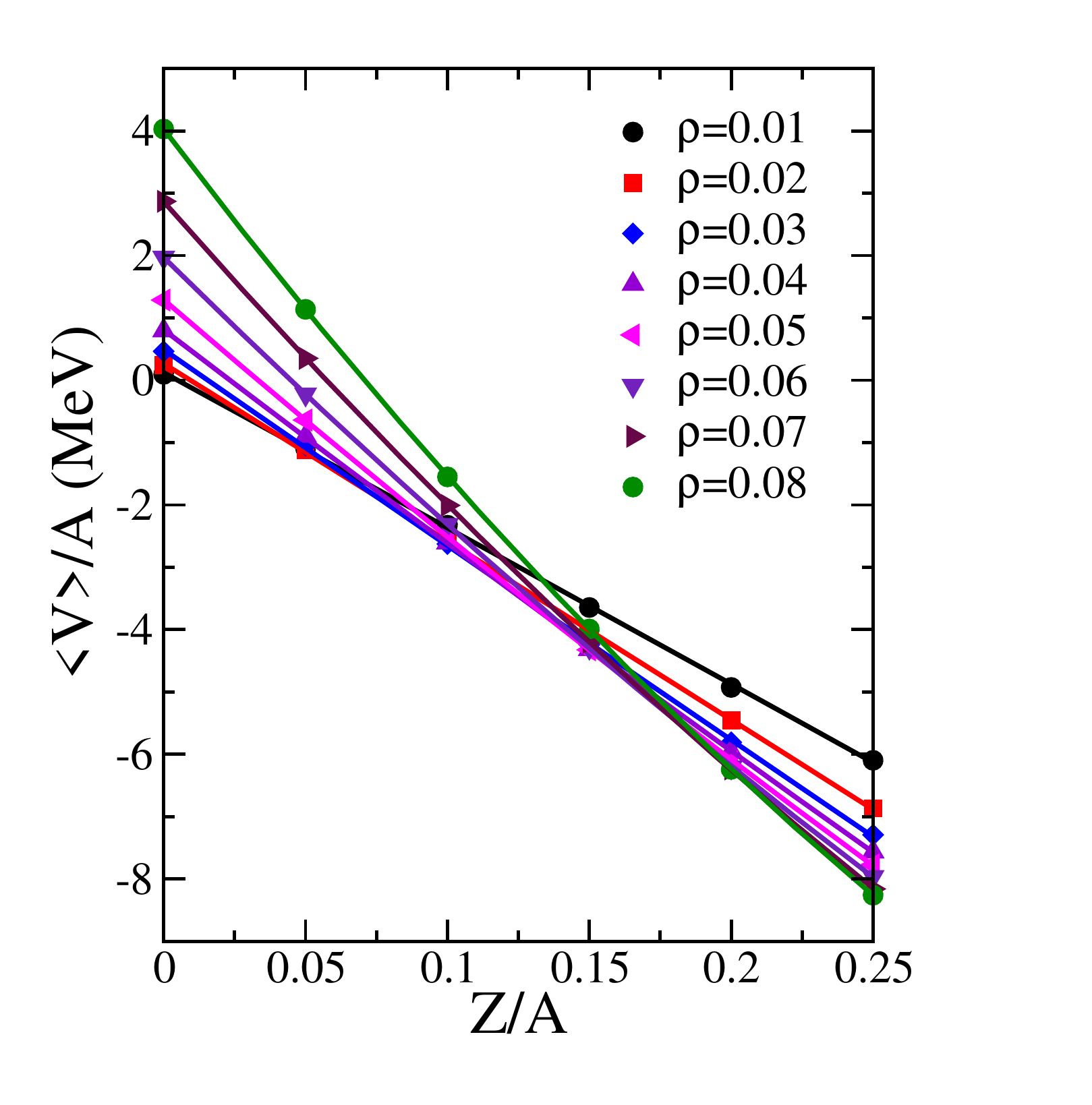}
\caption{(Color online) Monte-Carlo results for the total 
(nuclear-plus-Coulomb) potential energy of a system of 
$A\!=\!5,000$ nucleons as a function of density and proton 
fraction.}
\label{Fig2}
\end{figure}

The potential energy per nucleon $\langle V\rangle/A$ as a function of
density and proton fraction $y$ is displayed in Fig.~\ref{Fig2}. (Note
that we have used the symbol ``$y$'' to denote an arbitrary value for
the proton fraction; we reserve the symbol ``$Y_{p}$'' to denote the
value obtained from applying the condition of
$\beta$-equilibrium). The strong-nuclear interaction favors symmetric
$N\!=\!Z$ ({\sl i.e.,} $y\!=\!1/2$) clusters so $\langle V\rangle$ is
a monotonically decreasing function of $y$ (at least until
$y\!\simeq\!1/2$).  The $y\!=\!0$ values are related to the equation
of state of pure neutron matter and display the density dependence
predicted by the present model. To enforce $\beta$-equilibrium it is
convenient---as well as accurate---to fit the $y$ dependence of the
potential to a quadratic form. That is,
\begin{equation}
 \frac{\langle V\rangle(\rho,y)}{A} = v_{0}(\rho) + 
  v_{1}(\rho)y + v_{2}(\rho)y^{2} \;.
 \label{VFit} 
\end{equation}
The best-fit values obtained for the three coefficients are displayed
in Table~\ref{Table1}.

\begin{widetext}
\begin{center}
\begin{table}[h]
\begin{tabular}{|c|c|c|c|}
  \hline
   $\rho ({\rm fm}^{-3}) $ & $v_{0} $(MeV) 
   & $v_{1} $(MeV)  & $v_{2} $(MeV)  \\
  \hline
  \hline
    0.010 & 0.1221	& -24.8220 & -0.7801 \\
    0.020 & 0.2675	& -28.3990 & -0.8496 \\
    0.030 & 0.4923	& -31.5810 &   1.3126 \\
    0.040 & 0.8103	& -34.7880 &   5.0355 \\
    0.050 & 1.2924	& -39.2340 & 11.6480 \\
    0.060 & 1.9713	& -44.9300 & 20.6400 \\
    0.070 & 2.8642	& -51.6810 & 30.5050 \\
    0.080 & 4.0295	& -60.0800 & 43.6780 \\
\hline
\end{tabular}
 \caption{Quadratic fit to the $y$-dependence of $\langle V\rangle/A$ 
  as given in Eq.~(\ref{VFit}).}
\label{Table1}
\end{table}
\end{center}
\end{widetext}

To determine the optimal proton fraction we compute the total energy
per nucleon of the system ($E/A$) as a function of both density and
proton fraction $y$. The optimal proton fraction ($Y_{p}$) is then
obtained by minimizing $E/A$ with respect to $y$ at fixed density. We
write the energy per nucleon of the system in the following form:
\begin{equation}
  \frac{E}{A}-m_{n} = -y\Delta m + \frac{T_{n}}{A} 
   + y\frac{(T_{p}+T_{e})}{Z} + \frac{\langle V\rangle}{A}
   \approx
   \frac{T_{n}}{A} +  y\frac{T_{e}}{Z} + \frac{\langle V\rangle}{A}\;.
\end{equation}
Here $\Delta m\!\equiv\!m_{n}\!-\!m_{p}\!-\!m_{e}$ is the
neutron-proton-electron mass difference and $T_{n}$, $T_{p}$, and
$T_{e}$ represent the kinetic energy of the three constituents.
In the above equation the {\sl ``approximate''} sign has
been used to indicate that the contribution from $\Delta m$ is
negligible (see Fig.~\ref{Fig4}) and that the
electronic kinetic energy $T_{e}$---assumed to be that of a
relativistic electron gas---is much larger than the classical proton
contribution ($3yT/2$).  Note that the energy ({\sl kinetic plus rest
mass}) per particle of a relativistic Fermi gas of particles of mass
$m$ and number density $\rho\!=\!k_{\rm F}^{3}/3\pi^{2}$ may be
expressed in terms of the dimensionless Fermi momentum $x_{\rm
F}\!\equiv\!k_{\rm F}/m$ (and $y_{\rm F}\!\equiv\!\sqrt{1+x_{\rm
F}^{2}}$) in the following compact form:
\begin{equation}
   {\varepsilon}_{{}_{\rm FG}}(x_{\rm F}) = \frac{3m}{8} 
   \left[\frac{x_{\rm F}y_{\rm F}(x_{\rm F}^{2}+y_{\rm F}^{2})-\ln(x_{\rm F}+y_{\rm F})}
   {x_{\rm F}^{3}} \right] =
  \begin{cases} 
    \displaystyle{m + \frac{3k_{\rm F}^{2}}{10m} + \ldots}
    & \text{if $k_{\rm F}\!\ll\!m$,} \\ & \\
   \displaystyle{\frac{3}{4}k_{\rm F} + \frac{3m^{2}}{4k_{\rm F}} + \ldots}
    & \text{if $m\!\ll\!k_{\rm F}$.}
 \end{cases}
 \label{EFermiGas}
\end{equation}
\begin{figure}[htb]
\vspace{-0.05in}
\includegraphics[width=0.5\columnwidth,angle=0]{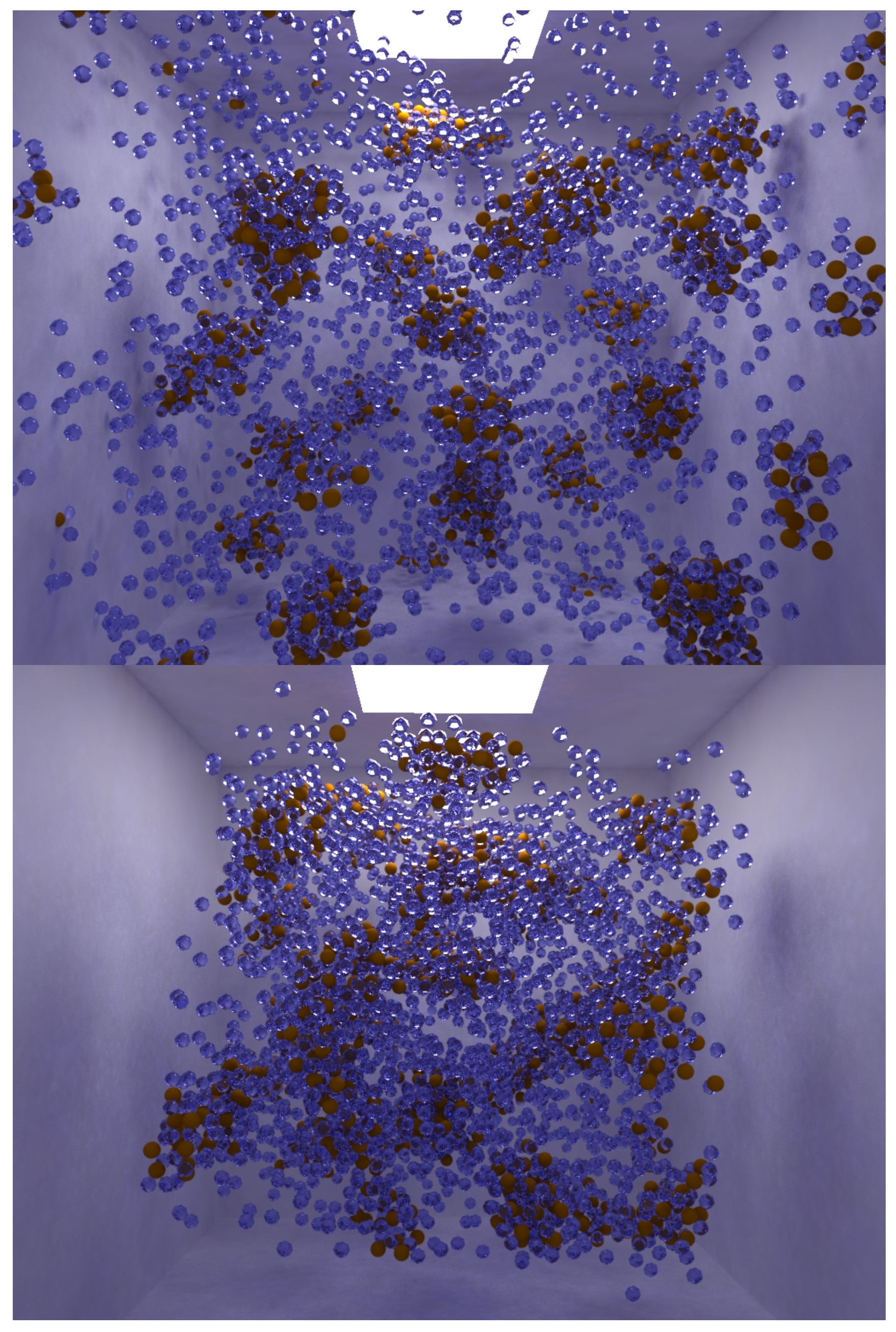}
\caption{(Color online) Two snapshots from a Monte Carlo 
simulation for a system of $A\!=\!4,000$ nucleons at a 
baryon density of $\rho\!=\!0.01~{\rm fm}^{-3}$ (upper panel) 
and $\rho\!=\!0.025~{\rm fm}^{-3}$ (lower panel). In both cases
a proton fraction of $Z/A\!=\!0.2$ and a temperature of 
$T\!=\!1$~MeV were used~\cite{Piekarewicz:2009zz}.}
\label{Fig3}
\end{figure}
Unfortunately, the neutron contribution $T_{n}$ to the energy of the
system is difficult to estimate. This is because at the densities of
relevance to the inner crust, the neutrons are either part of the
heavy clusters or part of the dilute neutron vapor. (See
Fig.~\ref{Fig3} for a snapshot of a Monte Carlo simulation that
clearly illustrates this scenario.) Presumably, the neutrons bound to
the heavy clusters are {\sl ``sluggish''} and thus may be treated
classically. The {\sl ``free''} neutrons on the other hand, should
probably be modeled as a quantum Fermi gas. However, deciding what
fraction of the neutrons are bound to clusters and what fraction
remains in the vapor is clearly a model-dependent question. Still, we
can make some progress in setting reasonable limits for the proton
fraction because the electronic Fermi energy dominates the kinetic
energy of the system. To set a lower limit we assume {\sl no}
contribution to the kinetic energy from the neutron vapor.
Conversely, a suitable upper limit for the proton fraction is obtained
by assuming that {\sl all} neutrons contribute to the kinetic energy.

\begin{figure}[htb]
\vspace{-0.05in}
\includegraphics[width=0.7\columnwidth,angle=0]{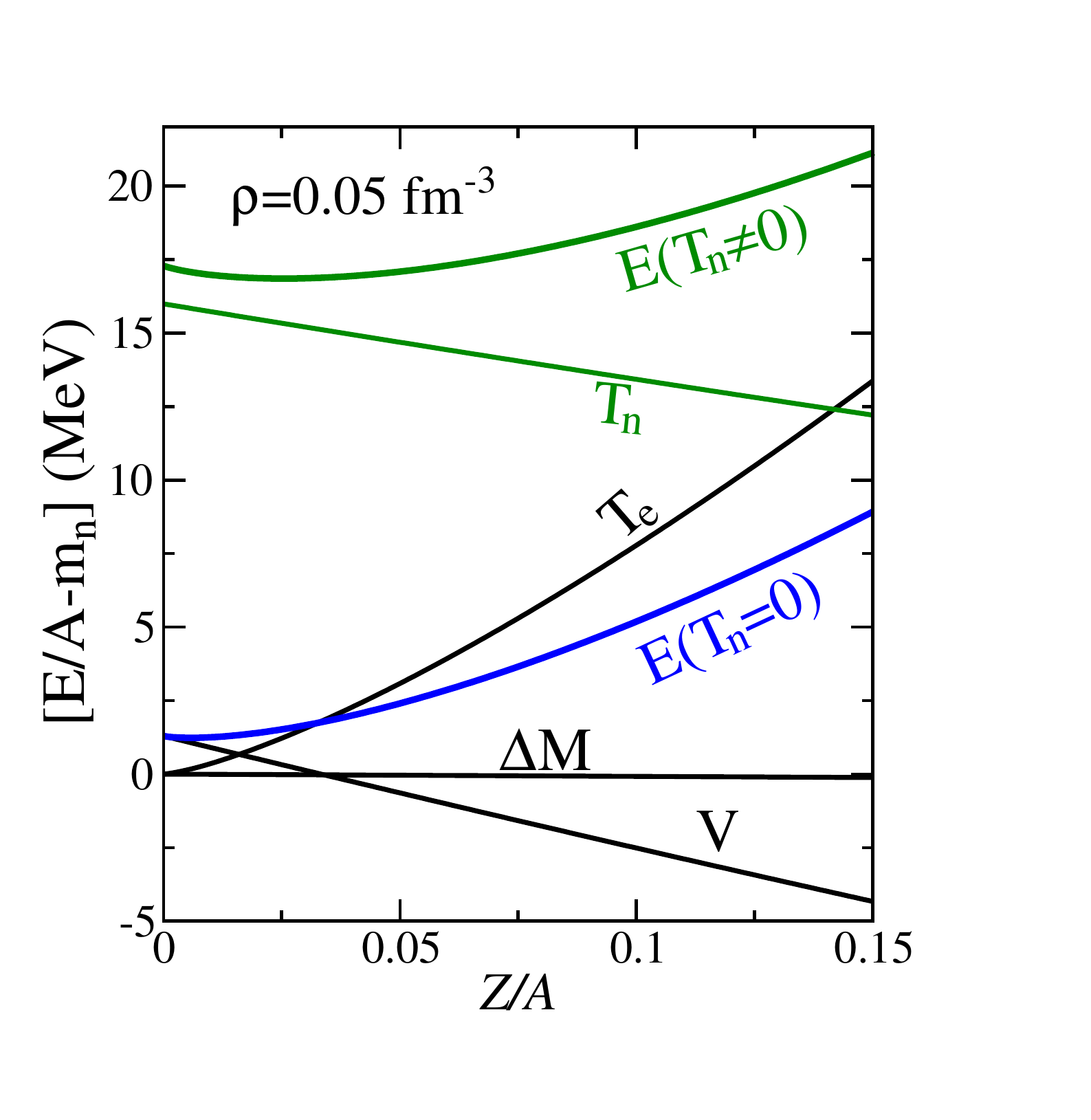}
\caption{(Color online) Enforcing beta equilibrium at a density
 of $\rho\!=\!0.05~{\rm fm}^{-3}$. The various contributions to
 the energy per nucleon are explained in the text.}
\label{Fig4}
\end{figure}

We illustrate how this process is implemented in Fig.~\ref{Fig4} where
the various contributions to the energy per nucleon have been plotted
as a function of the proton fraction for a fixed density of
$\rho\!=\!0.05~{\rm fm}^{-3}$. As alluded earlier, the contribution
from the mass difference $\Delta m$ is negligible. Hence, in the limit
of no contribution from the neutron vapor, the proton fraction emerges
from a competition between the potential energy---which favors
symmetric matter---and the electronic Fermi energy---which favors pure
neutron matter. Given that the electronic Fermi energy changes more
rapidly with proton fraction than $\langle V\rangle$, 
$\beta$-equilibrium is reached for the very small value of
$Y_{p}\!\simeq\!0.006$. As expected, adding the vapor contribution (as
a free Fermi gas of $N$ neutrons) increases the proton fraction by
almost a factor of 4; to $Y_{p}\!\simeq\!0.025$. Still, in both
cases the proton fraction is very small and significantly smaller---by
almost a factor of 10---than the $0.2$ value employed in our earlier
simulations~\cite{Horowitz:2004yf,Horowitz:2004pv,Horowitz:2005zb}.

\begin{figure}[htb]
\vspace{-0.05in}
\includegraphics[width=0.6\columnwidth,angle=0]{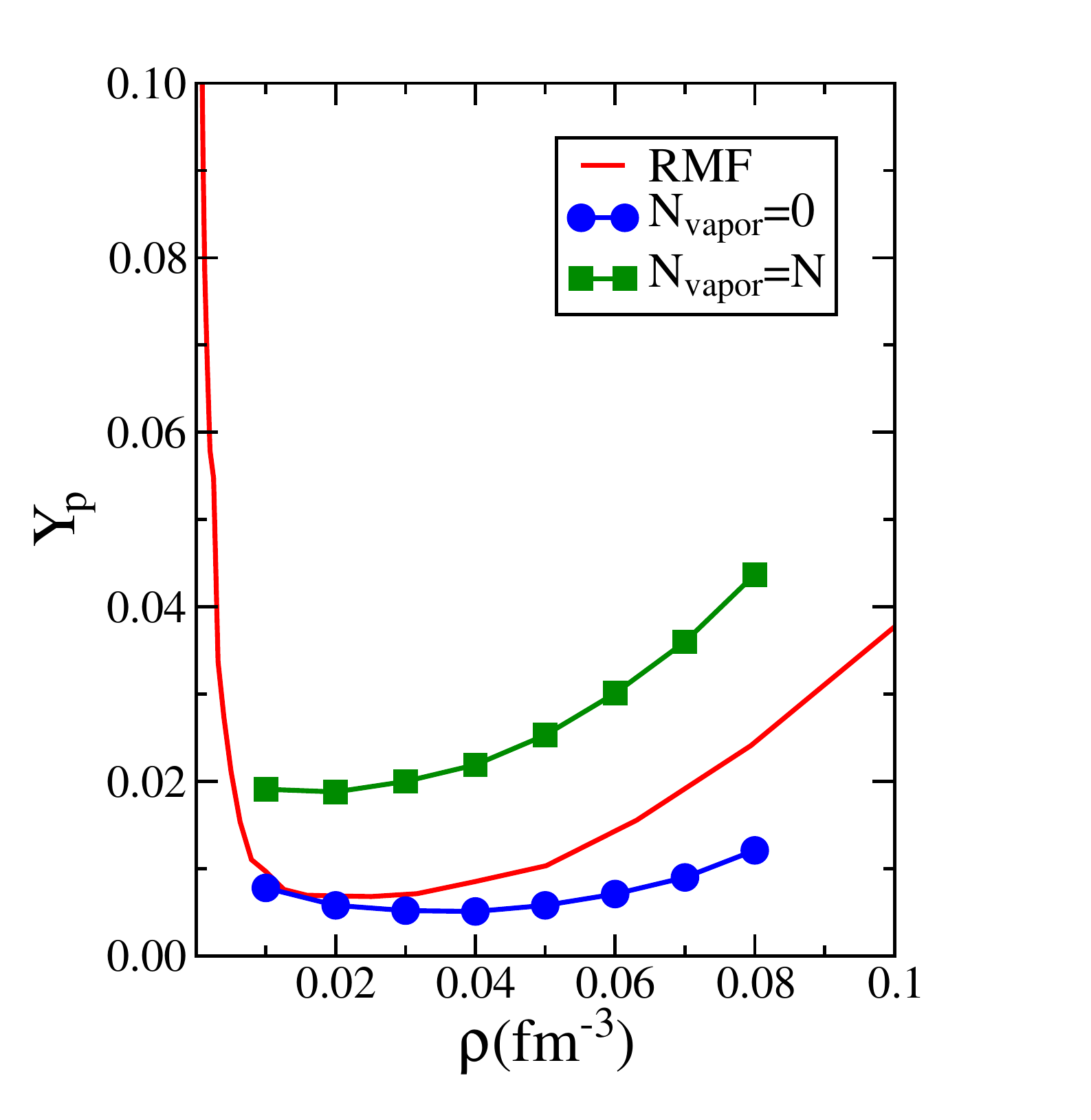}
\caption{(Color online) Proton fraction computed under the 
 assumption that either all the neutrons are bound into heavy clusters 
 $(N_{\rm vapor}=0)$ or all the neutrons contribute to the Fermi gas 
 energy $(N_{\rm vapor}\!=\!N)$ (see text for details). Also shown is
 a relativistic mean-field prediction for the proton fraction as 
 described in Ref.~\cite{Shen:2011kr}.}
\label{Fig5}
\end{figure}

Having implemented the above procedure at all densities, we display in
Fig.~\ref{Fig5} the predicted proton fraction as a function of baryon
density. Our results suggest very low proton fractions over the entire
range of densities explored in this work. This finding appears
consistent with predictions made using vastly different
approaches~\cite{Maruyama:2005vb,Avancini:2008zz,Avancini:2008kg,Shen:2011kr}.
For example, in Fig.~\ref{Fig5} results are also shown using the
relativistic mean-field (RMF) model of Ref.~\cite{Shen:2011kr}.  The
precipitous decline in the proton fraction displayed in the figure is
driven by the rapid increase with density of the electronic
contribution.  That is, at the densities of relevance to the inner
crust, electron capture is favored because the dilute neutron vapor
contributes little to the energy whereas the electrons contribute a
lot.  This represents robust physics that should be fairly model
independent. Thus, we must conclude that the proton fraction in
$\beta$-stable, neutron-star matter must be very small at the
densities of relevance to pasta formation.  Motivated by this finding,
we find imperative to determine whether pasta formation can occur 
in the proton-poor environment of the inner stellar crust. To do so, 
one must be able to identify observables that are particularly
sensitive to the formation of these exotic structures.  For reasons 
that will soon become clear, we focus on the static structure factor.

A fundamental physical observable that contains all information about
the excitations of the many-body system is the dynamic response
function $S({\bf k},\omega)$. The dynamic response function---which is
proportional to the scattering cross section---represents the
probability that the system be excited by a probe ({\sl e.g.,}
electron, neutrino, {\sl etc.}) that transfers momentum ${\bf k}$ and
energy $\omega$ into the system. To obtain dynamical information of
this kind one most resort to Molecular Dynamics simulations---which 
we hope to carry out in the near future. Unfortunately, Monte Carlo
simulations such as the ones implemented here can only reveal static
properties.  Yet the {\sl static structure factor} $S({\bf
k})$---obtained by integrating the dynamic response over
$\omega$---provides a particularly useful (static) observable that is
associated with the {\sl mean-square density fluctuations} in the
ground state~\cite{Fetter:1971}. Moreover, $S({\bf k})$ is intimately
related to a quantity that can be readily determined in computer
simulations: the {\sl pair-correlation} function $g({\bf r})$. Indeed,
$S({\bf k})$ and $g({\bf r})$ are simply Fourier transforms of each
other. That is~\cite{Vesely:2001},
\begin{equation}
   S({\bf k}) = 1 + \frac{N}{V} \int d^{3}r \Big(g({\bf r})-1\Big)
   {\Large{e}}^{-i{\bf k}\cdot{\bf r}} \;.
 \label{SofK}
\end{equation}
The pair-correlation function $g({\bf r})$ represents the probability 
of finding a pair of particles separated by a distance ${\bf r}$.  For 
a uniform fluid containing $N$ particles and confined to a simulation 
volume $V$, $g(r)$ (which is only a function of the magnitude of 
${\bf r}$) may be computed exclusively in terms of the 
instantaneous positions of the particles. That is,
\begin{equation}
   g(r) = 1 + \frac{V}{4\pi r^{2}N^{2}} 
   \Big\langle \sum_{i\ne j} \delta(r-|{\bf r}_{i}-{\bf r}_j|)\Big\rangle \;,
\label{GofR}
\end{equation}
where the ``{\sl brackets}'' represent an ensemble average.  Whereas
for a uniform fluid the one-body density is constant, interesting
two-body correlations emerge as a consequence of the inter-particle 
dynamics and/or quantum statistics. For example, the characteristic
short-range repulsion of the $NN$ interaction precludes particles 
from approaching each other. This results in a pair-correlation
function that vanishes at short separations (see Fig.~\ref{Fig6}).

Given that the static structure factor accounts for the {\sl
mean-square density fluctuations} in the ground state, it becomes a
particularly useful indicator of the critical behavior associated with
phase transitions---which themselves are characterized by the
development of large ({\sl i.e.,} macroscopic) fluctuations. Indeed,
the spectacular phenomenon of {\sl ``critical opalescence''} in fluids
is the macroscopic manifestation of abnormally large density
fluctuations---and thus abnormally large light scattering---near a
phase transition~\cite{Pathria:1996}. In this regard, the static
structure factor at {\sl zero-momentum transfer} provides a unique
connection to the thermodynamics of the
system~\cite{Pathria:1996}. That is,
\begin{equation}
   S({\bf k}\!=\!0) = 1 + \frac{N}{V} \int d^{3}r \Big(g({\bf r})-1\Big)
   = \frac{\langle N^{2}\rangle -\langle N\rangle^{2}}{\langle N\rangle} 
   =  \frac{k_{B}T}{\langle N\rangle}
       \left(\frac{\partial\langle N\rangle}{\partial\mu}\right)_{T,V}
   = \rho k_{B}T\kappa_{T} \;,
\label{SofKZero}
\end{equation}
where $k_{B}$ is Boltzmann's constant, $\mu$ is the chemical
potential, and $\kappa_{T}$ is the isothermal compressibility of the
system.  The isothermal compressibility $\kappa_{T}$ is reminiscent 
of the specific heat which accounts for energy, rather than density,
fluctuations. They both play a fundamental role in identifying the
onset of critical phenomena.

\begin{figure}[htb]
\vspace{-0.05in}
\includegraphics[width=0.7\columnwidth,angle=0]{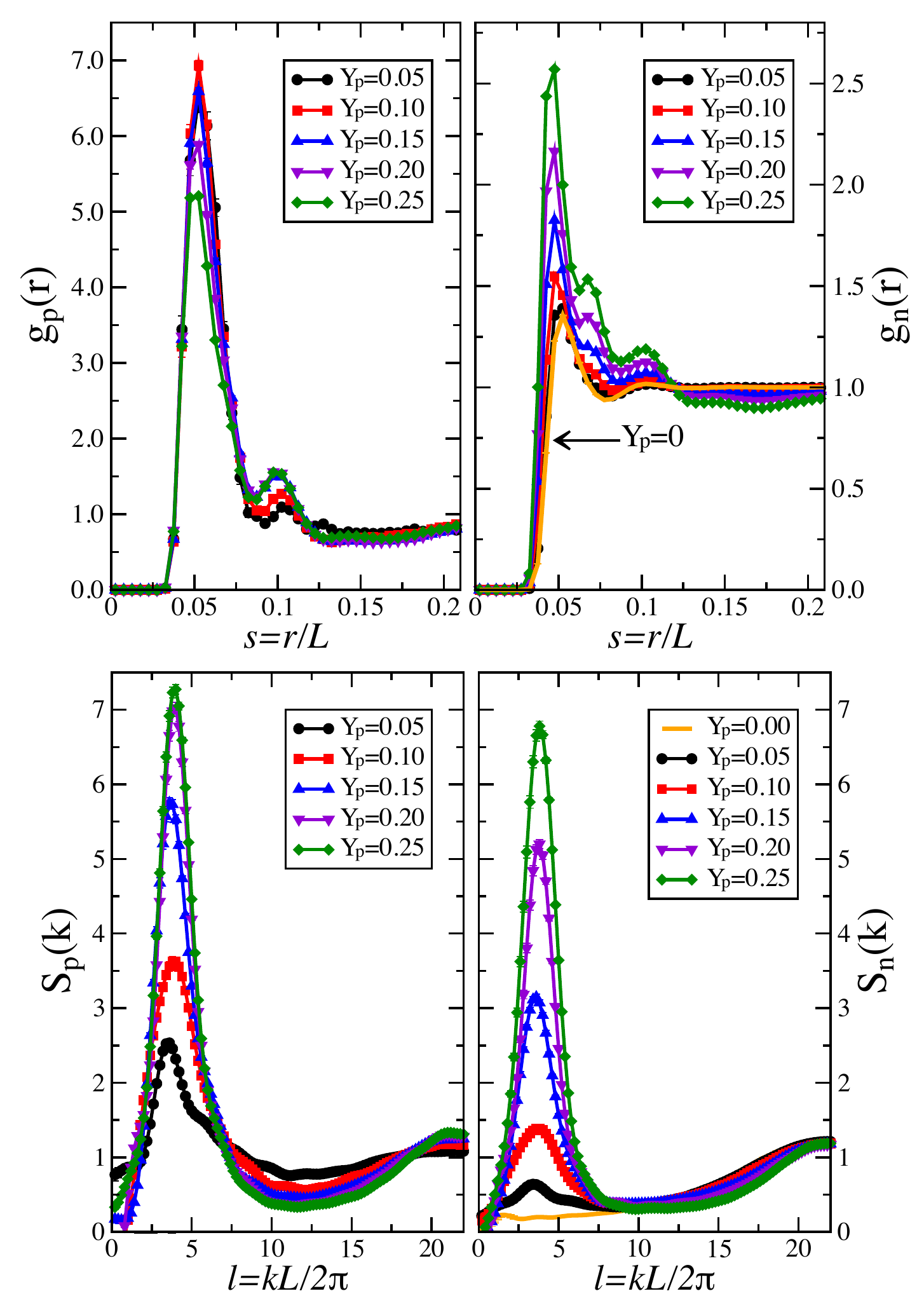}
\caption{(Color online) Pair correlation function $g(r)$ and static 
structure factor $S(k)$ for protons (left panels) and neutrons (right 
panels) for a variety of proton fractions at a fixed baryon density 
of $\rho\!=\!0.03~{\rm fm}^{-3}$.}
\label{Fig6}
\end{figure}

We conclude this section by displaying in Fig.~\ref{Fig6} the pair
correlation function $g(r)$ (upper panels) and static structure factor
$S(k)$ (lower panels) for both protons (left panels) and neutrons
(right panels) at the fixed density of $\rho\!=\!0.03~{\rm fm}^{-3}$. 
At this value of the density pasta structures are known to be 
present---at least for moderately large proton
fractions (see Fig.~\ref{Fig3}).  As expected, both $g_{n}(r)$ and
$g_{p}(r)$ vanish at small separations due to the characteristic
short-range repulsion of the $NN$ interaction.  Also in both cases,
the first peak is particularly large given that nearest neighbors are
strongly correlated within each cluster (see Fig.~\ref{Fig3}).
However, we also see qualitative differences between them.  In the
case of the neutrons, $g_{n}(r)$ displays the characteristic
oscillatory structure of a fluid that is associated with the presence
of nearest neighbors, next-to-nearest neighbors and so on.  Unlike 
$g_{n}(r)$, however, the proton correlation function $g_{p}(r)$ displays
only two prominent peaks---one large and one small.  This is
due to the absence of a proton vapor, as all protons are contained
within neutron-rich clusters. Whereas the first peak in $g_{p}(r)$
develops due to the strong correlation between protons in the 
same cluster, the second (significantly smaller) peak represents the
presence of protons in the neighboring clusters. This second peak
appears to grow appreciably with proton fraction but then
saturates. Yet it is unclear at this time whether the initial increase 
and eventual saturation may be significant. We will continue to 
explore this issue in the near future.

The two lower panels in Fig.~\ref{Fig6} display the static structure
factors obtained as the Fourier transform of the corresponding
pair-correlation functions. The static structure factor for the case
of pure neutron matter ($Y_{p}\!=\!0$) is shown for reference and
remains essentially structureless for all values of $k$. In contrast,
$S(k)$ for neutron-rich matter displays a prominent peak that becomes
progressively higher with increasing proton fraction.  The maximum in
$S(k)$ reflects the most prominent oscillatory structure in $g(r)$
which, in turn, is associated with the spatial (two-body) correlations
in the system. Alternatively, the maximum in $S(k)$ occurs at that
momentum transfer for which the probe ({\sl e.g.,} electrons) can most
efficiently scatter from the density fluctuations in the system. As in
the case of the second maximum in $g_{p}(r)$, here too the maximum in
$S_{p}(k)$ grows with proton fraction and then appears to saturate.
Whether this is significant in characterizing the emergence of the
nuclear pasta remains to be investigated. What is clear, however, is
that the static structure factor at zero momentum transfer
$S(k\!=\!0)$ does not show the required enhancement associated with
the onset of a phase transition.

\section{Conclusions}
\label{Conclusions}

Uniform neutron-rich matter at sub-saturation density is unstable
against cluster formation. At densities below the neutron-drip line
the ground-state of neutron-rich matter consists of a Coulomb crystal
of spherical (neutron-rich) nuclei embedded in a uniform sea of
electrons. It has been speculated that the transition from the
highly-ordered crystal to the uniform Fermi liquid is mediated by an
exotic phase known as {\sl ``nuclear pasta''}. The nuclear-pasta
phase---believed to exist in the inner crust of neutron stars---is
characterized by the emergence of exotic nuclear shapes of various
topologies that coexist with a dilute vapor of neutrons and
electrons. Critical to the formation of the pasta is the long-range
Coulomb interaction among protons, as it promotes the deformation of
the clusters and the development of exotic topological structures (see
Fig.~\ref{Fig3}).  Given this essential fact, it is important to
establish whether the proton fraction in the inner stellar crust is
large enough to favor pasta formation.

To compute the proton fraction in the inner stellar crust we relied on
Monte Carlo simulations assuming that nucleons interact via a
two-body, short-range nuclear interaction and a long-range Coulomb
repulsion. We employed a simple $NN$ interaction identical to the one
introduced in Ref.~\cite{Horowitz:2004yf} that was (approximately)
fitted to a few ground-state properties of finite nuclei, the
saturation properties of symmetric nuclear matter, and that precludes
the binding of pure neutron matter. Yet we improved on the work of
Ref.~\cite{Horowitz:2004yf} by treating the long-range Coulomb
interaction exactly via an Ewald summation.  Monte Carlo simulations
were performed for a system of $A\!=\!5,000$ nucleons at a variety of
densities and proton fractions (note that the simulations in
Ref.~\cite{Horowitz:2004yf} were limited to the single proton fraction
of $Y_{p}\!=\!0.2$). By doing so, the optimum proton fraction (at each
density) was determined by imposing $\beta$-equilibrium. The optimum
proton fraction emerges from a delicate competition between the
nuclear symmetry energy---which favors proton fraction of
$Y_{p}\!\lesssim\!1/2$---and the Fermi energy of the electrons---which
favors $Y_{p}\!=\!0$. For the model employed here, the rapid increase
of the electronic contribution with proton fraction dominates over the
nuclear contribution. This leads to {\sl very small proton fractions
at all densities of relevance to the inner crust}. For example, at a
density of $\rho\!=\!0.03~{\rm fm}^{-3}$ the proton fraction is
$Y_{p}\!\lesssim\!0.02$---at least a factor of 10 smaller than assumed
in Ref.~\cite{Horowitz:2004yf}. Such small values for the proton
fraction are consistent with those obtained using vastly different
approaches (see, for example,
Refs.~\cite{Maruyama:2005vb,Avancini:2008zz,Avancini:2008kg,Shen:2011kr}
and references therein). Ultimately, the small proton fractions
obtained in most approaches appears to be a direct consequence 
of the rapid increase with density of the electronic contribution.

Given these results and the predominant role played by the Coulomb
interaction in the formation of the nuclear pasta, it is only natural 
to ask whether such exotic structures can develop in the 
proton-poor environment of the inner stellar crust. To answer this
question it becomes critical to identify observables that may be
sensitive to pasta formation. Particularly useful in this case are
observables sensitive to fluctuations, as these tend to increase
rapidly near phase transitions. In this contribution we have focused
on the static structure factor $S(k)$ which is sensitive to density
fluctuations.  Moreover, the static structure factor is interesting as
it provides a natural meeting place for theory, experiment, and
computer simulations. Theoretically, $S(k)$ is the Fourier transform
of the pair-correlation function $g(r)$---a quantity that is amenable
to computer simulations. Experimentally, the proton (neutron) static
structure factor can actually be measured in electron (neutrino)
scattering experiments. In this work we have computed 
the pair-correlation function---for both protons and
neutrons---as a function of proton fraction for a fixed density of
$\rho\!=\!0.03~{\rm fm}^{-3}$.  In turn, the static structure factors
were obtained by performing the required Fourier transforms. Although
both $g(r)$ and $S(k)$ display interesting behavior that may be
indicative of significant structural changes in the system, we find no
clear evidence either in favor or against the formation of the nuclear
pasta.  In particular, the static structure factor at zero momentum
transfer $S(k\!=\!0)$ does not display the large enhancement
characteristic of a phase transition. Perhaps what is required to
clearly identify the onset and evolution of the pasta phase are
dynamical observables, such as various transport coefficients.  To do
so, one must rely on Molecular Dynamics (rather than Monte Carlo)
simulations. A calculation along these lines was reported in
Ref.~\cite{Horowitz:2008vf} where both the shear viscosity and thermal
conductivity were computed. Perhaps surprisingly, no dramatic increase
in the viscosity was observed as the system evolves from spherical to
exotic nuclear shapes. Clearly, more effort should---and will---be 
devoted along these lines.

In conclusion, our work revolved around three fundamental questions.
First, what are typical values for the proton fraction in the inner
stellar crust? Second, can exotic pasta structures develop in a
proton-deficient environment? Finally, which physical observables are
particularly sensitive to pasta formation?  We have found---as others
have before us using vastly different approaches---that the proton
fraction in the inner crust is very small indeed. However, at this
moment it is unclear whether exotic pasta structure can develop in
such proton-poor environments. In order to answer this question one
would have to identify physical observables that are sensitive to the
formation of the pasta. This task, however, has so far proven
elusive. In an effort to answer these open questions we plan to
perform Molecular Dynamics simulations with a very large number of
particles. By doing so we will be able to calculate both static and
dynamic observables. We trust that hidden among these host of
observables will be at least one that will respond dramatically to the
formation of the nuclear pasta.


\begin{acknowledgments}
 This work was supported in part by a grant from the U.S. Department
 of Energy DE-FD05-92ER40750. We thank C.J. Horowitz and G. Shen
 for useful discussions and for providing the RMF calculations
 presented in Fig.~\ref{Fig5}.
\end{acknowledgments}

\appendix*
\section{The Ewald Construction}

In this appendix we describe the derivation of the Ewald formula used
to compute the Coulomb potential of a system of $Z$ protons immersed
in a neutralizing and uniform electron background. Although such a
derivation is now part of the standard repertoire of Computer
Simulation textbooks (see for example
Refs.~\cite{Allen:1987,Frenkel:1996,Vesely:2001}) we present here a
detailed account in an attempt to keep the manuscript self-contained.

We consider a system of $Z$ point protons confined to a simulation cube 
of (finite) volume $V\!=\!L^{3}$ that is immersed in a neutralizing electron 
background of uniform density $\varrho_{e}=-eZ/V$. The Coulomb interaction 
for such a system is given by
\begin{equation}
  V_{Coul}({\bf r}_{1},\ldots,{\bf r}_{Z}) = 
 \frac{1}{2}\int_{V} \rho({\bf r}) \Phi({\bf r}) d^{3}r =
 \frac{1}{2}e\sum_{i=1}^{Z} \Phi({\bf r}_{i}) -
 \frac{eZ}{2V}\int_{V} \Phi({\bf r}) d^{3}r \;,
 \label{VCoul1}
\end{equation}
where $\Phi({\bf r})$ is the electrostatic potential that one must
compute as a solution of Poisson's equation. In an effort to minimize
{\sl finite-size} effects we employ periodic boundary conditions. That
is, the (finite) simulation box is exactly replicated an infinite
number of times---resulting in a perfect periodic tiling of all
space. It is this resulting periodic charge density that must be used
as the source of the electrostatic potential. In this periodic
approach, each proton within the simulation volume $V$ interacts not
only with the other $Z\!-\!1$ protons contained in the finite
simulation box but, in addition, with all the (infinitely many) {\sl
``images''} residing in the periodic boxes.  Clearly, such a
prescription appears highly impractical. In the case of short-range
interactions, such as the strong nucleon-nucleon interaction, periodic
boundary conditions pose no additional computational burden as each
individual proton in the simulation box interacts with only one---the
closest image---of the remaining $Z\!-\!1$ protons.  This is because
the next closest image must be at least half a box length ($L/2$)
away, thereby suppressing the contribution from all remaining
images provided, of course, that the range of the interaction is
significantly smaller than $L/2$.  However, the long-range Coulomb
interaction has no intrinsic scale so one must cope with the challenge
of dealing with all periodic images. Therein lies the power of the
{\sl Ewald summation}.

To compute the electrostatic potential that enters in
Eq.~(\ref{VCoul1}) one must solve Poisson's equation using the
periodic charge density (including image charges) as the source term.
That is,
\begin{equation}
 \nabla^{2}\Phi({\bf r}) = -4\pi\rho({\bf r}) \;,
 \label{Poisson}
\end{equation}
where
\begin{equation}
 \rho({\bf r}) = e\sum_{i=1}^{Z}\sum_{{\bf n}}
 \delta\Big({\bf r}-({\bf r}_{i}-{\bf n}L)\Big) - e\frac{Z}{V} \;.
\label{Rho0}
\end{equation}
Here ${\bf n}\!=\!(n_{x},n_{y},n_{z})$ is a {\sl translational vector} 
consisting of three arbitrary integers. To improve the convergence
properties of the (infinite) sum required to compute the Coulomb
interaction, Ewald introduced a distribution of {\sl smeared} charges
of magnitude $-e$ centered around the (point) protons. Indeed, 
Ewald's main {\sl ``trick''} consists in adding and subtracting the 
following charge density:
\begin{equation}
 \rho_{a}({\bf r}) = -e\sum_{i=1}^{Z}\sum_{{\bf n}}
 \delta_{a}\Big({\bf r}-({\bf r}_{i}-{\bf n}L)\Big) \;,
\label{Rhoa}
\end{equation}
where $\delta_{a}({\bf r})$ represents a diffuse charge distribution
(of unit charge) and smearing parameter $a$. As it is often done, 
we assume here a gaussian charge distribution of the form
\begin{equation}
 \delta_{a}({\bf r}) = \frac{1}{\left(\pi a^{2}\right)^{3/2}}
 \exp(-r^{2}/a^{2}) \;.
\label{Gaussian}
\end{equation}
By adding and subtracting the above smeared charge distribution
one may decompose the total charge density in two components;
a {\sl short-range} one and a {\sl long-range} one. That is,
\begin{subequations} 
\begin{align}
 & \rho({\bf r})  \equiv \varrho_{\rm sr}({\bf r}) + 
     \varrho_{\rm lr}({\bf r})\;, \quad {\rm where} \\
 & \varrho_{\rm sr}({\bf r}) =
  e\sum_{i=1}^{Z}\sum_{\bf n} \left[
 \delta\Big({\bf r}-({\bf r}_{i}-{\bf n}L)\Big) -
 \delta_{a} \Big({\bf r}-({\bf r}_{i}-{\bf n}L)\Big)\right] \;,
 \label{RhoSR} \\
 &\varrho_{\rm lr}({\bf r}) =e\sum_{i=1}^{Z}\sum_{\bf n}
  \delta_{a}\Big({\bf r}-({\bf r}_{i}-{\bf n}L)\Big) - e\frac{Z}{V}\;.
\label{RhoLR}
\end{align}
\end{subequations}

\subsection{Short-range electrostatic potential}
\label{SREPotential}

The short-range contribution to the electrostatic potential may be 
directly computed using Poisson's equation. In particular, the 
electrostatic potential due to the gaussian charge distribution 
given in Eq.~(\ref{Gaussian}) is given by
\begin{equation}
 \Phi_{a}({\bf r}) = \frac{{\rm erf}(r/a)}{r} \;,
\label{Phia}
\end{equation}
where ${\rm erf}(x)$ is the error function. As expected, the
electrostatic potential generated by such a smeared charge
distribution modifies the Coulomb potential at short distances
($r\!\lesssim\!a$) but then {\sl ``heels''} back rapidly to the
standard $1/r$ form as $r\!\gg\!a$. This results in an effective
screening of the bare proton charge for distances larger than 
the smearing parameter. Effectively, the long-range electrostatic
potential gets replaced by a short-range one. That is,
\begin{equation}
 \Phi({\bf r}) = \frac{1}{r}  \rightarrow
 \Phi_{\rm sr} ({\bf r}) = \frac{1}{r} - \frac{{\rm erf}(r/a)}{r}  
 \equiv \frac{{\rm erfc}(r/a)}{r} \;,
\label{PhiScreened}
\end{equation}
where ${\rm erfc}(x)\!=\!1\!-\!{\rm erf}(x)$ is the complement
of the error function. Using the principle of superposition, the
short-range component of the electrostatic potential is given
by
\begin{equation}
 \Phi_{\rm sr}({\bf r};a) = e\sum_{i=1}^{Z}\sum_{\bf n}  
 \frac{{\rm erfc}\Big[({\bf r}-{\bf r}_{i}+{\bf n}L)/a\Big]}
 {|{\bf r}-{\bf r}_{i}+{\bf n}L|} \;.
 \label{PhiSR}
\end{equation}

\subsection{Long-range electrostatic potential}
\label{LREPotential}

The remaining---long-range contribution---will be computed using
Fourier-transform techniques. Given that both the long-range component
of the charge density [$\varrho_{\rm lr}({\bf r})$ in
Eq.~(\ref{RhoLR})] and the resulting electrostatic potential are periodic
functions of ${\bf r}$, they can be expressed in terms of a Fourier
series as follows:
\begin{subequations} 
 \begin{align}
 & \Phi_{\rm lr}({\bf r}) = \frac{1}{V} \sum_{\bf k} 
     \Phi_{\rm lr}({\bf k}) \exp(i{\bf k}\cdot{\bf r})\;, 
 \label{FourierPhi0} \\  
 & \varrho_{\rm lr}({\bf r}) = \frac{1}{V} \sum_{\bf k} 
     \varrho_{\rm lr}({\bf k}) \exp(i{\bf k}\cdot{\bf r})\;
 \label{FourierRho0}
 \end{align}
\end{subequations}
where the allowed momenta ${\bf k}$ are quantized in units of
$2\pi/L$. In terms of the Fourier representations, Poisson's 
equation may be solved by inspection. That is,
\begin{equation}
 \Phi_{\rm lr}({\bf k})=\frac{4\pi}{k^{2}}\varrho_{\rm lr}({\bf k}) \;.
 \label{Phik}
\end{equation}
where the charge form factor $\varrho_{\rm lr}({\bf k})$ is given by
\begin{equation}
 \varrho_{\rm lr}({\bf k}) = 
 \int_{V} d^{3}r \exp(-i{\bf k}\cdot{\bf r}) 
  \left[e\sum_{i=1}^{Z}\sum_{\bf n}
  \delta_{a}({\bf r}-{\bf r}_{i}+{\bf n}L) - e\frac{Z}{V}\right]\;.
 \label{FourierRho1}
\end{equation}
Note that as it stands, $\Phi_{\rm lr}({\bf k})$ diverges at ${\bf
k}\!=\!0$ due to the long-range nature of the Coulomb interaction.  
As we will show below, the presence of a neutralizing electron background
leads to $\varrho_{\rm lr}({\bf k}\!=\!0)\!=\!0$ (as a long wavelength
probe can only resolve the total charge of the system)---effectively
removing this zero-momentum singularity.

One of the great simplifications that emerges from the periodicity of
the charge distribution is that the integral over the finite
simulation volume $V$ in Eq.~(\ref {FourierRho1}) can be transformed
into an integral over all of space ($\Omega$). This fact may now be
used to obtain the following simple and illuminating form for 
$\varrho_{\rm lr}({\bf k})$:
\begin{equation}
 \varrho_{\rm lr}({\bf k}) = e\sum_{i=1}^{Z}\int_{\Omega} 
   d^{3}r \exp(-i{\bf k}\cdot{\bf r}) \delta_{a}({\bf r}-{\bf r}_{i}) 
   - eZ\delta_{{\bf k},0}
  = F_{\rm ch}({\bf k}) \delta_{a}({\bf k})\Big(1-\delta_{{\bf k},0}\Big)\;,
 \label{FourierRho2}
\end{equation}
where $F_{\rm ch}({\bf k})$ is the charge form factor of the (point)
proton distribution and $\delta_{a}({\bf k})$ is the Fourier transform
of the smearing function. That is,
\begin{subequations} 
 \begin{align}
  & F_{\rm ch}({\bf k}) = e\sum_{i=1}^{Z} 
     \exp(-i{\bf k}\cdot{\bf r}_{i}) \;, 
     \label{ChargeFF} \\
  &\delta_{a}({\bf k}) = \int_{\Omega} d^{3}r 
     \exp(-i{\bf k}\cdot{\bf r})\delta_{a}({\bf r}) 
   =\exp(-k^{2}a^{2}/4) \;.
  \label{Gaussk}
\end{align}
\end{subequations}
Three features were instrumental in obtaining a form for $\varrho_{\rm
lr}({\bf k})$ that makes the formalism amenable to numerical
computations: (a) the periodicity of the charge distribution, (b) the
diffuseness of the single-proton density, and (c) the charge
neutrality of the system. First, the periodicity of the system enabled
one to transform an integral over the finite simulation volume $V$
over an integral over all of space, thereby simplifying
the calculation of $\delta_{a}({\bf k})$. Second, the
diffuseness of the single-proton density suppresses momentum
components that are much larger than $a^{-1}$, leading to a
rapid convergence of the momentum sum required to compute $\Phi_{\rm
lr}({\bf r})$ [see Eq.~(\ref{PhiLR})].  Finally, the neutralizing
electron background removes the zero-momentum component of
$\varrho_{\rm lr}({\bf k})$, rendering the electrostatic potential
finite at ${\bf k}\!=\!0$.

Using Eqs.~(\ref{FourierPhi0}),~(\ref{Phik}),~(\ref{FourierRho2}),
and~(\ref{Gaussk}), we obtain the following (numerically suitable)
form for the long-range component of the electrostatic potential:
\begin{equation} 
 \Phi_{\rm lr}({\bf r};a) = \frac{4\pi}{V} \sum_{{\bf k}\ne 0} 
  F_{\rm ch}({\bf k}) \frac{{\Large e}^{-k^{2}a^{2}/4}}{k^{2}} 
  e^{i{\bf k}\cdot{\bf r}} \;. 
\label{PhiLR}
\end{equation}

Having obtained the electrostatic potential using Ewald's
construction, we are now in a position to compute the resulting
Coulomb energy. As was done with the electrostatic potential, we
separate the Coulomb energy into a short-range and a long-range
component.

\subsection{Coulomb-Ewald potential}
\label{CEPotential}

The short-range component of the Coulomb potential is obtained by
inserting the short-range electrostatic potential $\Phi_{\rm sr}({\bf r})$ 
given in Eq.~(\ref{PhiSR}) into Eq.~(\ref{VCoul1}). We obtain
\begin{align}
 V_{\rm sr}({\bf r}_{1},\ldots,{\bf r}_{Z};a) &= 
 \frac{1}{2}e\sum_{i=1}^{Z} \Phi_{\rm sr}({\bf r}_{i}) -
 \frac{eZ}{2V}\int_{V} \Phi_{\rm sr}({\bf r}) d^{3}r 
  \nonumber \\ 
  &=\frac{1}{2}\sum_{i,j,{\bf n}}
  v_{\rm sr}({\bf r}_{i}-{\bf r}_{j}+{\bf n}L) -
 \frac{Z}{2V}\sum_{i,\bf{n}}
 \int_{V} v_{\rm sr}({\bf r}-{\bf r}_{i}+{\bf n}L) \,d^{3}r \;,
\label{VCoulSR1}
\end{align}
where we have introduced the modified---short-range---two-body
``Coulomb'' interaction as follows:
\begin{equation}
 v_{\rm sr}({\bf r})=e^{2}\,\frac{{\rm erfc}(r/a)}{r} \;.
\end{equation}
The electronic contribution [second term in Eq.~(\ref{VCoulSR1})] may
be evaluated by using (as we did earlier) the periodicity of the
problem to re-write the integral over the finite simulation volume as
an integral over all of space. That is,
\begin{equation}
    \sum_{i,\bf{n}}\int_{V} v_{\rm sr}({\bf r}-{\bf r}_{i}+{\bf n}L)\,d^{3}r 
 = \sum_{i}\int_{\Omega}v_{\rm sr}({\bf r}-{\bf r}_{i})\,d^{3}r  \;.
 \label{Electronic1}
\end{equation}
Given that the integral is now over all space, one then can shift the
variable of integration to remove all dependence on the proton
coordinates ${\bf r}_{i}$ to obtain a (constant) term proportional 
to $Z$:
\begin{equation}
  \sum_{i=1}^{Z}\int_{\Omega}v_{\rm sr}({\bf r}-{\bf r}_{i})\,d^{3}r  
  = Z \int_{\Omega}v_{\rm sr}({\bf r})\,d^{3}r 
  = Ze^{2}\pi a^{2} \;.
  \label{Electronic2}
\end{equation}
As it stands, the sum over proton coordinates in Eq.~(\ref{VCoulSR1})]
diverges because of spurious self-interactions ({\sl i.e.,} the $Z$
terms with $i\!=\!j$ and ${\bf n}\!=\!0$). To render this sum finite
we must remove all self-interactions among the point protons. That is,
\begin{align}
 V_{\rm self} = \frac{1}{2}Z \lim_{r\rightarrow 0} \frac{e^{2}}{r}
  &= \frac{Ze^{2}}{2} \lim_{r\rightarrow 0}
  \left[\frac{{\rm erfc}(r/a)}{r} + \frac{{\rm erf}(r/a)}{r}\right] 
  \nonumber \\
  & = \frac{1}{2}\sum_{i=1}^{Z} v_{\rm sr}({\bf r}_{i}-{\bf r}_{i}) 
     +  \frac{Ze^{2}}{a\sqrt{\pi}} \;.
\end{align}
By subtracting this self-interaction term from Eq.~(\ref{VCoulSR1}),
we obtain the short-range contribution to the electrostatic Coulomb
energy:
\begin{equation}
  V_{\rm sr}({\bf r}_{1},\ldots,{\bf r}_{Z};a) =
  \frac{1}{2}\sum_{i,j,{\bf n}}{}^{\!'}
   v_{\rm sr}({\bf r}_{ij}+{\bf n}L) -
   \frac{Ze^{2}}{a\sqrt{\pi}} -
   \frac{\pi a^{2}}{2} \frac{(Ze)^{2}}{V} \;,
\label{VCoulSR2}
\end{equation}
where ${\bf r}_{ij}\!\equiv\!{\bf r}_{i}\!-\!{\bf r}_{j}$ and the
prime is used to indicate that self-interaction terms (namely, the $Z$
terms with $i\!=\!j$ and ${\bf n}\!=\!0$) should all be omitted from
the sum.

As in the case of the short-range interaction, the long-range
component of the Coulomb potential is obtained by inserting 
the long-range electrostatic potential $\Phi_{\rm lr}({\bf r})$ 
[Eq.~(\ref{PhiLR})] into Eq.~(\ref{VCoul1}). We obtain
\begin{align}
  V_{\rm lr}({\bf r}_{1},\ldots,{\bf r}_{Z};a) &= 
 \frac{1}{2}e\sum_{i=1}^{Z} \Phi_{\rm lr}({\bf r}_{i}) -
 \frac{eZ}{2V}\int_{V} \Phi_{\rm lr}({\bf r}) d^{3}r 
  \nonumber \\ 
  &=\frac{2\pi}{V}\sum_{{\bf k}\ne 0}
  |F_{\rm ch}({\bf k})|^{2}\,\frac{{\Large e}^{-k^{2}a^{2}/4}}{k^{2}} 
  -\frac{eZ}{2V}\Phi_{\rm lr}({\bf k}\!=\!0) \;. 
\label{VCoulLR1}
\end{align}
The first term in the above equation is obtained by directly
substituting $\Phi_{\rm lr}({\bf r})$ into the proton sum and then
using the definition of the (point) charge form factor introduced in
Eq.~(\ref{ChargeFF}). The second term is proportional to the
zero-momentum component of the potential and vanishes because of the
overall charge neutrality of the system [see Eq.~(\ref{FourierRho2})].

Collecting both---short- and long-range---contributions we arrive 
at a form of the Coulomb potential that is both well defined and 
rapidly convergent: 
\begin{align}
  V_{\rm Coul}({\bf r}_{1},\ldots,{\bf r}_{Z}) &=
   \frac{2\pi}{V}\sum_{{\bf k}\ne 0}
   |F_{\rm ch}({\bf k})|^{2}\,\frac{{\Large e}^{-k^{2}a^{2}/4}}{k^{2}} +
  \frac{1}{2}\sum_{i,j,{\bf n}}{}^{\!'}
   v_{\rm sr}({\bf r}_{ij}+{\bf n}L) 
   \nonumber \\ & -
   \frac{Ze^{2}}{a\sqrt{\pi}} -
   \frac{\pi a^{2}}{2} \frac{(Ze)^{2}}{V} \;.
\label{VCoulEwald}
\end{align}

Before leaving the derivation of the Ewald method and proceed to test
its accuracy and convergence properties, we present the explicit form
of the Coulomb potential that will be used in our simulations. To do
so we take advantage of the fact that the long-range Coulomb potential
is an interaction with no intrinsic scale. To start, we use the
box-size $L$ to define dimensionless coordinates (${\bf s}$) and
momenta (${\bf l}$) as follows:
\begin{subequations}
\begin{align}
  {\bf r} & = L{\bf s} \;, \\ 
  {\bf k} & = \frac{2\pi}{L}{\bf l} \;.
\end{align}
\label{slDef}
\end{subequations}
By introducing these simple definitions into Eq.~(\ref{VCoulEwald}),
we isolate the {\sl dimensionfull} parameter $v_{0}\!\equiv\!e^{2}/L$ 
from the {\sl dimensionless} many-body Coulomb potential. That is,
\begin{equation}
  V_{\rm Coul}({\bf r}_{1},\ldots,{\bf r}_{Z}) = 
  \left(\frac{e^{2}}{L}\right)  
  U_{\rm Coul}({\bf s}_{1},\ldots,{\bf s}_{Z}) \;,
 \label{UDef}
\end{equation}
where
\begin{equation}
  U_{\rm Coul}({\bf s}_{1},\ldots,{\bf s}_{Z}) =
   \frac{1}{2}\sum_{{\bf l}\ne 0}
   |F_{\rm ch}({\bf l})|^{2}\,w(l) +
  \frac{1}{2}\sum_{i,j,{\bf n}}{}^{\!'}
   u_{\rm sr}({\bf s}_{ij}+{\bf n}) 
 + U'_{0} \;.
\label{VCoulEwald2}
\end{equation}
Here the long-range interaction is written in terms of  the 
(point-proton) charge form factor $F_{\rm ch}({\bf l})$ and 
the modified Coulomb propagator $w(l)$
\begin{subequations}
\begin{align}
   F_{\rm ch}({\bf l}) &= \sum_{j=1}^{Z} 
  \exp(-2\pi i{\bf l}\cdot{\bf s}_{j}) \;, \\
   w(l) &=\frac{\exp(-\pi^{2}s_{0}^{2}\,l^{2})}
  {\pi l^{2}}
  \quad ({\rm with} \; s_{0}\equiv a/L) \;, 
 \end{align}
 \label{FFandW}
\end{subequations}
the short-range interaction is given by
\begin{equation}
 u_{\rm sr}({\bf s})=\frac{{\rm erfc}(s/s_{0})}{s} \;,
 \label{uSRdef}
\end{equation}
and $U'_{0}$ is a constant:
\begin{equation}
 U'_{0} = - \frac{Z}{\sqrt{\pi}s_{0}} -
                \frac{\pi s_{0}^{2}Z^{2}}{2} \;.
  \label{CFF}
\end{equation}

Evidently, the Coulomb potential must be independent of the auxiliary
smearing parameter $s_{0}$. This is convenient as one may use this
independence to test the reliability of the method. Once such a test has
been performed, one can use this freedom by selecting a value of
$s_{0}$ that will expedite the computation without sacrificing
accuracy. In particular, in this contribution we will employ a
smearing parameter that is significantly smaller than the box length
({\sl i.e.,} $s_{0}\!\approx\!0.1$). This ensures that the {\sl
minimum image convention} may be used to compute the short-range part
of the Coulomb potential. To compute the long-range part of the
potential one must evaluate a sum over momentum modes. For a small
value of $s_{0}$, as the one adopted here, a very large number of modes
must be included in order to achieve convergence.  As a result, this
part of the computation can be quite expensive. Thus, one often
resorts to methods that handle the Fourier sum more efficiently, such
as {\sl particle-mesh} approaches~\cite{Frenkel:1996}. In this
contribution we implement a simple ``poor's-man'' approach that
requires {\sl pre-computing} the potential on a relatively fine
mesh. During the simulation, computing the long-range part of the
potential is then reduced to a simple interpolation scheme. To do so
we write the square of the of the charge form-factor appearing in
Eq.~(\ref{VCoulEwald2}) as follows:
\begin{equation}
   |F_{\rm ch}({\bf l})|^{2} = \sum_{i=1}^{Z} \sum_{j=1}^{Z} 
  \exp(-2\pi i{\bf l}\cdot{\bf s}_{ij}) = Z +
  \sum_{i\ne j}^{Z} \exp(-2\pi i{\bf l}\cdot{\bf s}_{ij}) \;.
\label{FF2}
\end{equation}
In this way the long-range component of the Coulomb potentials
reduces---as in the case of the short-range component---to a sum 
of two-body terms (plus a constant).  That is,
\begin{equation}
  U_{\rm lr}({\bf s}_{1},\ldots,{\bf s}_{Z}) \equiv
  \frac{1}{2}\sum_{{\bf l}\ne 0}
   |F_{\rm ch}({\bf l})|^{2}\,w(l) =
  \frac{1}{2}\sum_{i\ne j}
   u_{\rm lr}({\bf s}_{ij})+\frac{Z}{2}u_{\rm lr}(0) \;,
\label{UCoulLR1}
\end{equation}
where the (long-range) two-body interaction has been
defined by
\begin{equation}
 u_{\rm lr}({\bf s})=\sum_{{\bf l}\ne 0} 
  w(l) \exp(-2\pi i{\bf l}\cdot{\bf s}) \;.
 \label{uLRdef}
\end{equation}
Finally, the dimensionless Coulomb interaction to be used 
in the Monte Carlo simulations is given by the following
expression:
\begin{equation}
  U_{\rm Coul}({\bf s}_{1},\ldots,{\bf s}_{Z}) =
   \frac{1}{2}\sum_{i\ne j}
   \Big[u_{\rm lr}({\bf s}_{ij}) +
          u_{\rm sr}({\bf s}_{ij}) \Big] + U_{0} \;.
\label{VCoulEwald3}
\end{equation}
where
\begin{equation}
 U_{0} = \frac{Z}{2}u_{\rm lr}(0) 
          - \frac{Z}{\sqrt{\pi}s_{0}} 
          - \frac{\pi s_{0}^{2}Z^{2}}{2} \;.
 \label{ConstantU}
\end{equation}

The Ewald construction is a powerful method that may be tested in a
variety of ways.  First, one may compare against well known results for
the Coulomb energy of simple crystals. Second, whereas the Ewald
construction introduces a dependence on the smearing parameter $a$,
the {\sl ``bare''} Coulomb potential should be insensitive to it---as
long as $a$ is neither too big nor too small. We have implemented such
ideas in Fig.~\ref{FigA} where the Coulomb energy of a system
consisting of $Z$ protons localized to the sites of a simple cubic
lattice and immersed in a uniform electron background has been
computed. The figure displays the competition between the long-range
(repulsive) and the short-range (attractive) contributions to the
Coulomb energy as a function of the smearing parameter $a$ and number
of protons. The energy is given in units of $e^{2}/a_{\rm latt}$,
where $a_{\rm latt}\!=\!\rho_{Z}^{1/3}$ is the lattice constant and
$\rho_{Z}$ is the proton density. We obtain for all values of $a$ and
all system sizes an energy per particle of $-2.837297479/2$, where
$2.837297479$ is the Madelung constant for a simple-cubic 
crystal~\cite{Brush:1966}.  Remarkably, the Ewald method can already
reproduce the exact answer (up to nine significant figures!) for a
system with two protons per side.

\begin{figure}[htb]
\vspace{-0.05in}
\includegraphics[width=0.6\columnwidth,angle=0]{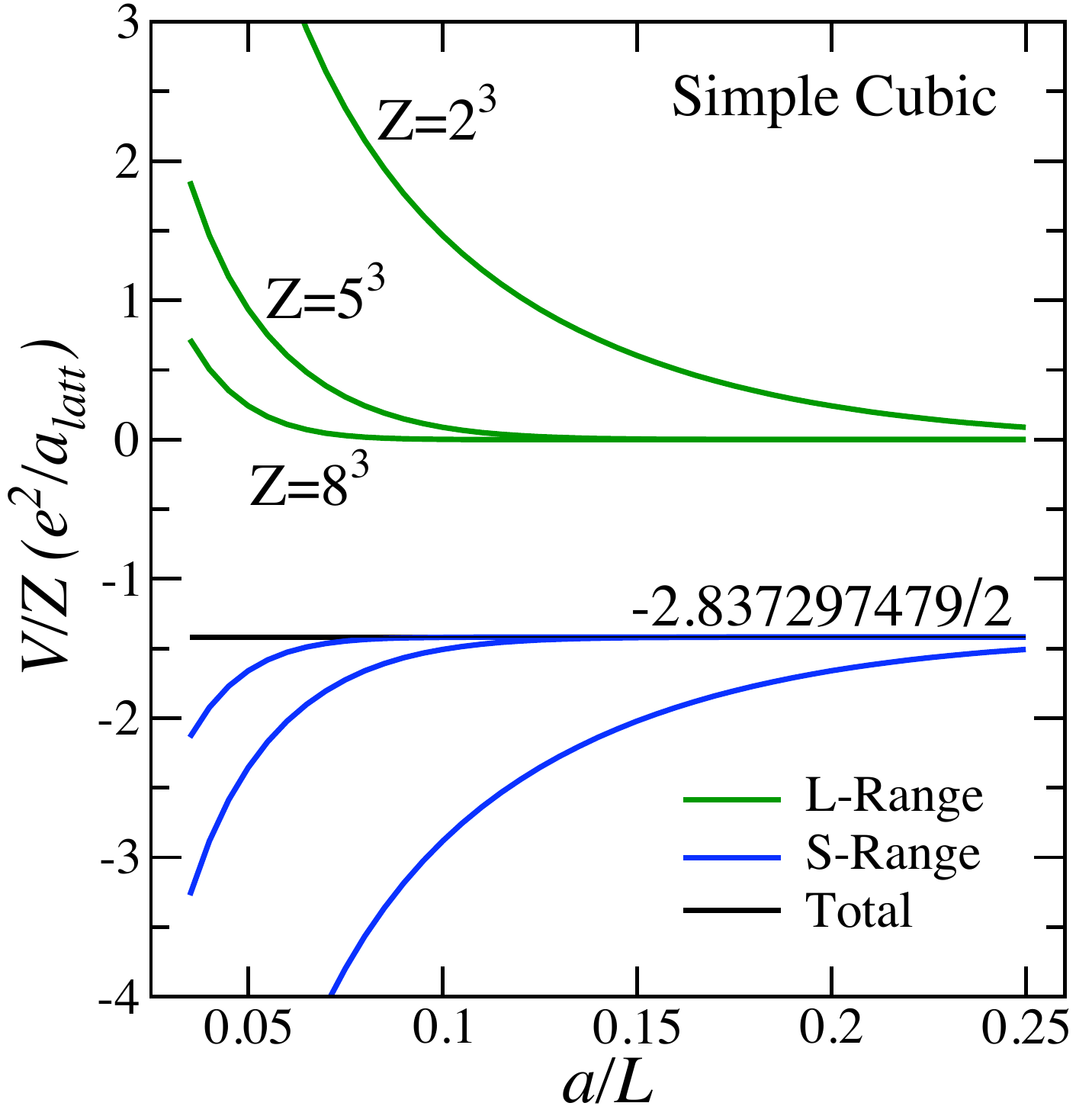}
\caption{(Color online) Coulomb energy per particle together with
its long- and short-range contributions for a simple cubic lattice 
of $Z$ protons immersed in a neutralizing and uniform electron
background. The Coulomb energy is given in units of 
$e^{2}/a_{\rm latt}$ ($a_{\rm latt}$ is the lattice constant) and the 
dimensionless Madelung constant for a simple cubic lattice is 
equal to $2.837297479$~\cite{Brush:1966}.}
\label{FigA}
\end{figure}

\vfill\eject
\bibliography{../../ReferencesJP.bib}

\begin{thebibliography}{31}
\expandafter\ifx\csname natexlab\endcsname\relax\def\natexlab#1{#1}\fi
\expandafter\ifx\csname bibnamefont\endcsname\relax
  \def\bibnamefont#1{#1}\fi
\expandafter\ifx\csname bibfnamefont\endcsname\relax
  \def\bibfnamefont#1{#1}\fi
\expandafter\ifx\csname citenamefont\endcsname\relax
  \def\citenamefont#1{#1}\fi
\expandafter\ifx\csname url\endcsname\relax
  \def\url#1{\texttt{#1}}\fi
\expandafter\ifx\csname urlprefix\endcsname\relax\def\urlprefix{URL }\fi
\providecommand{\bibinfo}[2]{#2}
\providecommand{\eprint}[2][]{\url{#2}}

\bibitem[{\citenamefont{Weber}(1999)}]{Weber:1999}
\bibinfo{author}{\bibfnamefont{F.}~\bibnamefont{Weber}},
  \emph{\bibinfo{title}{Pulsars as Astrophysical Laboratories for Nuclear and
  Particle Physics}} (\bibinfo{publisher}{Institute of Physics Publishing},
  \bibinfo{address}{Bristol, UK}, \bibinfo{year}{1999}).

\bibitem[{\citenamefont{Glendenning}(2000)}]{Glendenning:2000}
\bibinfo{author}{\bibfnamefont{N.~K.} \bibnamefont{Glendenning}},
  \emph{\bibinfo{title}{Compact Stars}} (\bibinfo{publisher}{Springer-Verlag
  New York}, \bibinfo{year}{2000}).

\bibitem[{\citenamefont{Piekarewicz}(2009)}]{Piekarewicz:2009zz}
\bibinfo{author}{\bibfnamefont{J.}~\bibnamefont{Piekarewicz}},
  \bibinfo{journal}{AIP Conf. Proc.} \textbf{\bibinfo{volume}{1182}},
  \bibinfo{pages}{937} (\bibinfo{year}{2009}).

\bibitem[{\citenamefont{Horowitz and Piekarewicz}(2001)}]{Horowitz:2000xj}
\bibinfo{author}{\bibfnamefont{C.~J.} \bibnamefont{Horowitz}} \bibnamefont{and}
  \bibinfo{author}{\bibfnamefont{J.}~\bibnamefont{Piekarewicz}},
  \bibinfo{journal}{Phys. Rev. Lett.} \textbf{\bibinfo{volume}{86}},
  \bibinfo{pages}{5647} (\bibinfo{year}{2001}).

\bibitem[{\citenamefont{Baym et~al.}(1971)\citenamefont{Baym, Pethick, and
  Sutherland}}]{Baym:1971pw}
\bibinfo{author}{\bibfnamefont{G.}~\bibnamefont{Baym}},
  \bibinfo{author}{\bibfnamefont{C.}~\bibnamefont{Pethick}}, \bibnamefont{and}
  \bibinfo{author}{\bibfnamefont{P.}~\bibnamefont{Sutherland}},
  \bibinfo{journal}{Astrophys. J.} \textbf{\bibinfo{volume}{170}},
  \bibinfo{pages}{299} (\bibinfo{year}{1971}).

\bibitem[{\citenamefont{Ruester et~al.}(2006)\citenamefont{Ruester, Hempel, and
  Schaffner-Bielich}}]{Ruester:2005fm}
\bibinfo{author}{\bibfnamefont{S.~B.} \bibnamefont{Ruester}},
  \bibinfo{author}{\bibfnamefont{M.}~\bibnamefont{Hempel}}, \bibnamefont{and}
  \bibinfo{author}{\bibfnamefont{J.}~\bibnamefont{Schaffner-Bielich}},
  \bibinfo{journal}{Phys. Rev.} \textbf{\bibinfo{volume}{C73}},
  \bibinfo{pages}{035804} (\bibinfo{year}{2006}).

\bibitem[{\citenamefont{Roca-Maza and Piekarewicz}(2008)}]{RocaMaza:2008ja}
\bibinfo{author}{\bibfnamefont{X.}~\bibnamefont{Roca-Maza}} \bibnamefont{and}
  \bibinfo{author}{\bibfnamefont{J.}~\bibnamefont{Piekarewicz}},
  \bibinfo{journal}{Phys. Rev.} \textbf{\bibinfo{volume}{C78}},
  \bibinfo{pages}{025807} (\bibinfo{year}{2008}).

\bibitem[{\citenamefont{Ravenhall et~al.}(1983)\citenamefont{Ravenhall,
  Pethick, and Wilson}}]{Ravenhall:1983uh}
\bibinfo{author}{\bibfnamefont{D.~G.} \bibnamefont{Ravenhall}},
  \bibinfo{author}{\bibfnamefont{C.~J.} \bibnamefont{Pethick}},
  \bibnamefont{and} \bibinfo{author}{\bibfnamefont{J.~R.}
  \bibnamefont{Wilson}}, \bibinfo{journal}{Phys. Rev. Lett.}
  \textbf{\bibinfo{volume}{50}}, \bibinfo{pages}{2066} (\bibinfo{year}{1983}).

\bibitem[{\citenamefont{Hashimoto et~al.}(1984)\citenamefont{Hashimoto, Seki,
  and Yamada}}]{Hashimoto:1984}
\bibinfo{author}{\bibfnamefont{M.}~\bibnamefont{Hashimoto}},
  \bibinfo{author}{\bibfnamefont{H.}~\bibnamefont{Seki}}, \bibnamefont{and}
  \bibinfo{author}{\bibfnamefont{M.}~\bibnamefont{Yamada}},
  \bibinfo{journal}{Prog. Theor. Phys.} \textbf{\bibinfo{volume}{71}},
  \bibinfo{pages}{320} (\bibinfo{year}{1984}).

\bibitem[{\citenamefont{Fetter and Walecka}(1971)}]{Fetter:1971}
\bibinfo{author}{\bibfnamefont{A.~L.} \bibnamefont{Fetter}} \bibnamefont{and}
  \bibinfo{author}{\bibfnamefont{J.~D.} \bibnamefont{Walecka}},
  \emph{\bibinfo{title}{Quantum Theory of Many Particle Systems}}
  (\bibinfo{publisher}{McGraw-Hill, New York}, \bibinfo{year}{1971}).

\bibitem[{\citenamefont{Jamei et~al.}(2005)\citenamefont{Jamei, Kivelson, and
  Spivak}}]{Jamei:2005}
\bibinfo{author}{\bibfnamefont{R.}~\bibnamefont{Jamei}},
  \bibinfo{author}{\bibfnamefont{S.}~\bibnamefont{Kivelson}}, \bibnamefont{and}
  \bibinfo{author}{\bibfnamefont{B.}~\bibnamefont{Spivak}},
  \bibinfo{journal}{Phys. Rev. Lett.} \textbf{\bibinfo{volume}{94}},
  \bibinfo{eid}{056805} (\bibinfo{year}{2005}).

\bibitem[{\citenamefont{Oyamatsu and Iida}(2007)}]{Oyamatsu:2006vd}
\bibinfo{author}{\bibfnamefont{K.}~\bibnamefont{Oyamatsu}} \bibnamefont{and}
  \bibinfo{author}{\bibfnamefont{K.}~\bibnamefont{Iida}},
  \bibinfo{journal}{Phys. Rev.} \textbf{\bibinfo{volume}{C75}},
  \bibinfo{pages}{015801} (\bibinfo{year}{2007}).

\bibitem[{\citenamefont{Horowitz
  et~al.}(2004{\natexlab{a}})\citenamefont{Horowitz, Perez-Garcia, and
  Piekarewicz}}]{Horowitz:2004yf}
\bibinfo{author}{\bibfnamefont{C.~J.} \bibnamefont{Horowitz}},
  \bibinfo{author}{\bibfnamefont{M.~A.} \bibnamefont{Perez-Garcia}},
  \bibnamefont{and}
  \bibinfo{author}{\bibfnamefont{J.}~\bibnamefont{Piekarewicz}},
  \bibinfo{journal}{Phys. Rev.} \textbf{\bibinfo{volume}{C69}},
  \bibinfo{pages}{045804} (\bibinfo{year}{2004}{\natexlab{a}}).

\bibitem[{\citenamefont{Horowitz
  et~al.}(2004{\natexlab{b}})\citenamefont{Horowitz, Perez-Garcia, Carriere,
  Berry, and Piekarewicz}}]{Horowitz:2004pv}
\bibinfo{author}{\bibfnamefont{C.~J.} \bibnamefont{Horowitz}},
  \bibinfo{author}{\bibfnamefont{M.~A.} \bibnamefont{Perez-Garcia}},
  \bibinfo{author}{\bibfnamefont{J.}~\bibnamefont{Carriere}},
  \bibinfo{author}{\bibfnamefont{D.~K.} \bibnamefont{Berry}}, \bibnamefont{and}
  \bibinfo{author}{\bibfnamefont{J.}~\bibnamefont{Piekarewicz}},
  \bibinfo{journal}{Phys. Rev.} \textbf{\bibinfo{volume}{C70}},
  \bibinfo{pages}{065806} (\bibinfo{year}{2004}{\natexlab{b}}).

\bibitem[{\citenamefont{Horowitz et~al.}(2005)\citenamefont{Horowitz,
  Perez-Garcia, Berry, and Piekarewicz}}]{Horowitz:2005zb}
\bibinfo{author}{\bibfnamefont{C.~J.} \bibnamefont{Horowitz}},
  \bibinfo{author}{\bibfnamefont{M.~A.} \bibnamefont{Perez-Garcia}},
  \bibinfo{author}{\bibfnamefont{D.~K.} \bibnamefont{Berry}}, \bibnamefont{and}
  \bibinfo{author}{\bibfnamefont{J.}~\bibnamefont{Piekarewicz}},
  \bibinfo{journal}{Phys. Rev.} \textbf{\bibinfo{volume}{C72}},
  \bibinfo{pages}{035801} (\bibinfo{year}{2005}).

\bibitem[{\citenamefont{Watanabe et~al.}(2003)\citenamefont{Watanabe, Sato,
  Yasuoka, and Ebisuzaki}}]{Watanabe:2003xu}
\bibinfo{author}{\bibfnamefont{G.}~\bibnamefont{Watanabe}},
  \bibinfo{author}{\bibfnamefont{K.}~\bibnamefont{Sato}},
  \bibinfo{author}{\bibfnamefont{K.}~\bibnamefont{Yasuoka}}, \bibnamefont{and}
  \bibinfo{author}{\bibfnamefont{T.}~\bibnamefont{Ebisuzaki}},
  \bibinfo{journal}{Phys. Rev.} \textbf{\bibinfo{volume}{C68}},
  \bibinfo{pages}{035806} (\bibinfo{year}{2003}).

\bibitem[{\citenamefont{Watanabe et~al.}(2005)\citenamefont{Watanabe, Maruyama,
  Sato, Yasuoka, and Ebisuzaki}}]{Watanabe:2004tr}
\bibinfo{author}{\bibfnamefont{G.}~\bibnamefont{Watanabe}},
  \bibinfo{author}{\bibfnamefont{T.}~\bibnamefont{Maruyama}},
  \bibinfo{author}{\bibfnamefont{K.}~\bibnamefont{Sato}},
  \bibinfo{author}{\bibfnamefont{K.}~\bibnamefont{Yasuoka}}, \bibnamefont{and}
  \bibinfo{author}{\bibfnamefont{T.}~\bibnamefont{Ebisuzaki}},
  \bibinfo{journal}{Phys. Rev. Lett.} \textbf{\bibinfo{volume}{94}},
  \bibinfo{pages}{031101} (\bibinfo{year}{2005}).

\bibitem[{\citenamefont{Watanabe et~al.}(2009)\citenamefont{Watanabe, Sonoda,
  Maruyama, Sato, Yasuoka et~al.}}]{Watanabe:2009vi}
\bibinfo{author}{\bibfnamefont{G.}~\bibnamefont{Watanabe}},
  \bibinfo{author}{\bibfnamefont{H.}~\bibnamefont{Sonoda}},
  \bibinfo{author}{\bibfnamefont{T.}~\bibnamefont{Maruyama}},
  \bibinfo{author}{\bibfnamefont{K.}~\bibnamefont{Sato}},
  \bibinfo{author}{\bibfnamefont{K.}~\bibnamefont{Yasuoka}},
  \bibnamefont{et~al.}, \bibinfo{journal}{Phys. Rev. Lett.}
  \textbf{\bibinfo{volume}{103}}, \bibinfo{pages}{121101}
  (\bibinfo{year}{2009}).

\bibitem[{\citenamefont{Maruyama et~al.}(2005)\citenamefont{Maruyama, Tatsumi,
  Voskresensky, Tanigawa, and Chiba}}]{Maruyama:2005vb}
\bibinfo{author}{\bibfnamefont{T.}~\bibnamefont{Maruyama}},
  \bibinfo{author}{\bibfnamefont{T.}~\bibnamefont{Tatsumi}},
  \bibinfo{author}{\bibfnamefont{D.~N.} \bibnamefont{Voskresensky}},
  \bibinfo{author}{\bibfnamefont{T.}~\bibnamefont{Tanigawa}}, \bibnamefont{and}
  \bibinfo{author}{\bibfnamefont{S.}~\bibnamefont{Chiba}},
  \bibinfo{journal}{Phys. Rev.} \textbf{\bibinfo{volume}{C72}},
  \bibinfo{pages}{015802} (\bibinfo{year}{2005}).

\bibitem[{\citenamefont{Avancini et~al.}(2008)\citenamefont{Avancini, Menezes,
  Alloy, Marinelli, Moraes et~al.}}]{Avancini:2008zz}
\bibinfo{author}{\bibfnamefont{S.}~\bibnamefont{Avancini}},
  \bibinfo{author}{\bibfnamefont{D.}~\bibnamefont{Menezes}},
  \bibinfo{author}{\bibfnamefont{M.}~\bibnamefont{Alloy}},
  \bibinfo{author}{\bibfnamefont{J.}~\bibnamefont{Marinelli}},
  \bibinfo{author}{\bibfnamefont{M.}~\bibnamefont{Moraes}},
  \bibnamefont{et~al.}, \bibinfo{journal}{Phys. Rev.}
  \textbf{\bibinfo{volume}{C78}}, \bibinfo{pages}{015802}
  (\bibinfo{year}{2008}).

\bibitem[{\citenamefont{Avancini et~al.}(2009)\citenamefont{Avancini, Brito,
  Marinelli, Menezes, de~Moraes et~al.}}]{Avancini:2008kg}
\bibinfo{author}{\bibfnamefont{S.}~\bibnamefont{Avancini}},
  \bibinfo{author}{\bibfnamefont{L.}~\bibnamefont{Brito}},
  \bibinfo{author}{\bibfnamefont{J.}~\bibnamefont{Marinelli}},
  \bibinfo{author}{\bibfnamefont{D.}~\bibnamefont{Menezes}},
  \bibinfo{author}{\bibfnamefont{M.}~\bibnamefont{de~Moraes}},
  \bibnamefont{et~al.}, \bibinfo{journal}{Phys.Rev.}
  \textbf{\bibinfo{volume}{C79}}, \bibinfo{pages}{035804}
  (\bibinfo{year}{2009}).

\bibitem[{\citenamefont{Newton and Stone}(2009)}]{Newton:2009zz}
\bibinfo{author}{\bibfnamefont{W.}~\bibnamefont{Newton}} \bibnamefont{and}
  \bibinfo{author}{\bibfnamefont{J.}~\bibnamefont{Stone}},
  \bibinfo{journal}{Phys. Rev.} \textbf{\bibinfo{volume}{C79}},
  \bibinfo{pages}{055801} (\bibinfo{year}{2009}).

\bibitem[{\citenamefont{Shen et~al.}(2011)\citenamefont{Shen, Horowitz, and
  Teige}}]{Shen:2011kr}
\bibinfo{author}{\bibfnamefont{G.}~\bibnamefont{Shen}},
  \bibinfo{author}{\bibfnamefont{C.}~\bibnamefont{Horowitz}}, \bibnamefont{and}
  \bibinfo{author}{\bibfnamefont{S.}~\bibnamefont{Teige}},
  \bibinfo{journal}{Phys. Rev.} \textbf{\bibinfo{volume}{C83}},
  \bibinfo{pages}{035802} (\bibinfo{year}{2011}).

\bibitem[{\citenamefont{Allen and Tildesley}(1987)}]{Allen:1987}
\bibinfo{author}{\bibfnamefont{M.~P.} \bibnamefont{Allen}} \bibnamefont{and}
  \bibinfo{author}{\bibfnamefont{D.~J.} \bibnamefont{Tildesley}},
  \emph{\bibinfo{title}{Computer Simulation of Liquids}}
  (\bibinfo{publisher}{Oxford University Press Inc., New York},
  \bibinfo{year}{1987}).

\bibitem[{\citenamefont{Frenkel and Smit}(1996)}]{Frenkel:1996}
\bibinfo{author}{\bibfnamefont{D.}~\bibnamefont{Frenkel}} \bibnamefont{and}
  \bibinfo{author}{\bibfnamefont{B.}~\bibnamefont{Smit}},
  \emph{\bibinfo{title}{Understanding Molecular Simulations}}
  (\bibinfo{publisher}{Academic Press, San Diego}, \bibinfo{year}{1996}).

\bibitem[{\citenamefont{Vesely}(2001)}]{Vesely:2001}
\bibinfo{author}{\bibfnamefont{F.~J.} \bibnamefont{Vesely}},
  \emph{\bibinfo{title}{Computational Physics: An Introduction}}
  (\bibinfo{publisher}{Kluwer Academic}, \bibinfo{address}{New York},
  \bibinfo{year}{2001}).

\bibitem[{\citenamefont{Ewald}(1921)}]{Ewald:1921}
\bibinfo{author}{\bibfnamefont{P.~P.} \bibnamefont{Ewald}},
  \bibinfo{journal}{Ann. Phys.} \textbf{\bibinfo{volume}{369}},
  \bibinfo{pages}{253} (\bibinfo{year}{1921}).

\bibitem[{\citenamefont{Toukmaji and Board}(1996)}]{Toukmaji_199673}
\bibinfo{author}{\bibfnamefont{A.~Y.} \bibnamefont{Toukmaji}} \bibnamefont{and}
  \bibinfo{author}{\bibfnamefont{J.~A.} \bibnamefont{Board}},
  \bibinfo{journal}{Computer Physics Communications}
  \textbf{\bibinfo{volume}{95}}, \bibinfo{pages}{73 } (\bibinfo{year}{1996}).

\bibitem[{\citenamefont{Pathria}(1996)}]{Pathria:1996}
\bibinfo{author}{\bibfnamefont{R.~K.} \bibnamefont{Pathria}},
  \emph{\bibinfo{title}{Statistical Mechanics}}
  (\bibinfo{publisher}{Butterworth-Heinemann}, \bibinfo{address}{Oxford},
  \bibinfo{year}{1996}), \bibinfo{edition}{2nd} ed.

\bibitem[{\citenamefont{Horowitz and Berry}(2008)}]{Horowitz:2008vf}
\bibinfo{author}{\bibfnamefont{C.~J.} \bibnamefont{Horowitz}} \bibnamefont{and}
  \bibinfo{author}{\bibfnamefont{D.~K.} \bibnamefont{Berry}},
  \bibinfo{journal}{Phys. Rev.} \textbf{\bibinfo{volume}{C78}},
  \bibinfo{pages}{035806} (\bibinfo{year}{2008}).

\bibitem[{\citenamefont{Brush et~al.}(1966)\citenamefont{Brush, Sahlin, and
  Teller}}]{Brush:1966}
\bibinfo{author}{\bibfnamefont{S.~G.} \bibnamefont{Brush}},
  \bibinfo{author}{\bibfnamefont{H.~L.} \bibnamefont{Sahlin}},
  \bibnamefont{and} \bibinfo{author}{\bibfnamefont{E.~J.}
  \bibnamefont{Teller}}, \bibinfo{journal}{J. Chem. Phys.}
  \textbf{\bibinfo{volume}{45}}, \bibinfo{pages}{2102} (\bibinfo{year}{1966}).

\end{thebibliography}
\end{document}